\documentclass[11pt]{article} 
\pdfoutput=1

\usepackage{amsmath}
\usepackage[norelsize]{algorithm2e}
\usepackage{scalerel}
\usepackage{tikz-cd}
\usepackage{ulem}
\usepackage{float}
\restylefloat{figure}
\usepackage{stackengine}
\usepackage[english]{babel}
\usepackage[light,condensed,math]{anttor}
\usepackage[T1]{fontenc}
\usepackage{kpfonts}
\usepackage{comment}
\usepackage{array}
\usepackage{lmodern}
\usepackage[mathscr]{euscript}

\usepackage{breqn}
\usepackage{todonotes}
\usepackage{slashed}
\usepackage{youngtab}
\usepackage{amsmath}
\usepackage{amssymb}
\usepackage{amsxtra}
\usepackage{color}
\usepackage[margin=1.2in]{geometry}

\newcommand{\dir}{\slashed{D}}
\newcommand{\ve}{{\varepsilon}}

\newcommand{\BC}{{\mathbb{C}}}

\newcommand{\BP}{{\mathbb{P}}}

\newcommand{\BR}{{\mathbb{R}}}

\newcommand {\ac} {\mathfrak{a}}

\def\beq{\begin{equation}}
\def\eeq{\end{equation}}

\newcommand\tr{{\rm tr}}

\newcommand{\mM}{{\mathfrak M}}

\newcommand{\ba} {\mathbf{a}}

\newcommand {\BT}   {\mathbb T}

\newcommand {\bN}   {\mathbf{N}}
\newcommand {\bM}   {\mathbf{M}}

\newcommand {\bQ}   {\mathbf{Q}}
\newcommand {\bR}   {\mathbf{R}}

\newcommand {\qe} {\mathfrak q}

\newcommand {\bnu}{ {\boldsymbol{\nu}}}

\newcommand {\bS}{ \mathbf{S}}

\newcommand {\BZ}   {\mathbb Z}

\newcommand{\bY}{\mathbf{Y}}

\newcommand {\CalA} {\mathcal A}

\newcommand {\CalE} {\mathcal E}

\newcommand {\CalG} {\mathcal G}
\newcommand {\CalH} {\mathcal H}
\newcommand {\CalI} {\mathcal I}

\newcommand {\CalL} {\mathcal L}
\newcommand {\CalM} {\mathcal M}
\newcommand {\CalN} {\mathcal N}
\newcommand {\CalO} {\mathcal O}
\newcommand {\CalP} {\mathcal P}
\newcommand {\CalQ} {\mathcal Q}
\newcommand {\CalR} {\mathcal R}

\newcommand {\CalV} {\mathcal V}
\newcommand {\CalX} {\mathcal X}

\newcommand {\CalW} {\mathcal W}
\newcommand {\CalZ} {\mathcal Z}

\newcommand {\zb} {{\bar z}}

\newcommand {\wb} {{\bar w}}

\newcommand{\fo}{\vert\kern -.03in\_}

\newcommand {\ii} {\mathrm{i}}
\newcommand{\Tr}{\text{Tr}}
\newcommand{\pa}{{\partial}}

\newcommand {\4}{\underline{\bf 4}}
\newcommand {\6}{\underline{\bf 6}}
\usepackage{float}
\restylefloat{figure}

\newcommand{\boxit}[1]{\vbox{\hrule\hbox{\vrule\kern8pt
\vbox{\hbox{\kern8pt}\hbox{\vbox{#1}}\hbox{\kern8pt}}
\kern8pt\vrule}\hrule}}
\newcommand{\mathboxit}[1]{\vbox{\hrule\hbox{\vrule\kern8pt\vbox{\kern8pt
\hbox{$\displaystyle #1$}\kern8pt}\kern8pt\vrule}\hrule}}

\newcommand{\Vg}{\mathsf{V}_{\gamma}}

\newcommand{\Eg}{\mathsf{E}_{\gamma}}

\title{\textsc{BPS/CFT correspondence IV: \\ sigma models and defects in gauge theory}}
\author{Nikita Nekrasov\footnote{Simons Center for Geometry and
    Physics, Stony Brook University, Stony Brook, NY 11794\, , 
    \newline{\tiny on leave of absence from:} 
    Kharkevich IITP RAS, Moscow, ITEP, Moscow .\newline
    e-mail: nnekrasov@scgp.stonybrook.edu}}
\date{}

\begin{document}
\maketitle

\begin{abstract}
 
 Quantum field theory $L_1$ on spacetime $X_{1}$ can be coupled to another quantum field theory $L_2$ on a spacetime $X_{2}$ via the third quantum field theory $L_{12}$ living on $X_{12} = X_{1} \cap X_{2}$. 
We explore several such constructions with two and four dimensional $X_{1}, X_{2}$'s and zero and two dimensional $X_{12}$'s, in the context of ${\CalN}=2$ supersymmetry, non-perturbative Dyson-Schwinger equations, and BPS/CFT correspondence. The companion paper \cite{N6} will show that the BPZ and KZ equations of two dimensional conformal field theory are obeyed by the half-BPS surface defects
in quiver ${\CalN}=2$ gauge theories.  

\end{abstract}

\section{Introduction}

Quantum field theory, with all its successes in describing the world of elementary particles, still lacks proper mathematically rigorous definition. One of the actively discussed topics is the locality of the space of states of the theory, how entangled are the states, and, more generally, the spatial distribution of the elementary building blocks of the finite energy states of the theory. 
In recent years the concept of generalized symmetries, acting between different quantum field theories, has been emerging. A possible way of realizing such symmetries is by merging two quantum field theories across a defect, as discussed in \cite{NSt} and, less explicitly, in the last section of \cite{NS1}. 

Given two Euclidean Lagrangian quantum field theories: one with some fields ${\bf\Phi}_{1}$ defined on a spacetime manifold $X_{1}$ and the action $L_{1}({\bf\Phi}_{1})$, and another with some fields ${\bf\Phi}_{2}$ defined on a spacetime manifold $X_{2}$ and the action $L_{2}({\bf\Phi}_{2})$, we can couple them along the intersection $X_{12} = X_{1} \cap X_{2}$ assuming it is non-empty. For example, suppose another set of fields ${\bf\Psi}$ live on $X_{12}$ only, with the Lagrangian $L_{12}({\bf\Psi} ; {\bf\Phi}_{1} \vert_{X_{12}} , {\bf\Phi}_{2} \vert_{X_{12}})$ containing the couplings to the restrictions of the ``bulk'' fields onto the intersection. 

\bigskip

\centerline{\includegraphics[width=5cm]{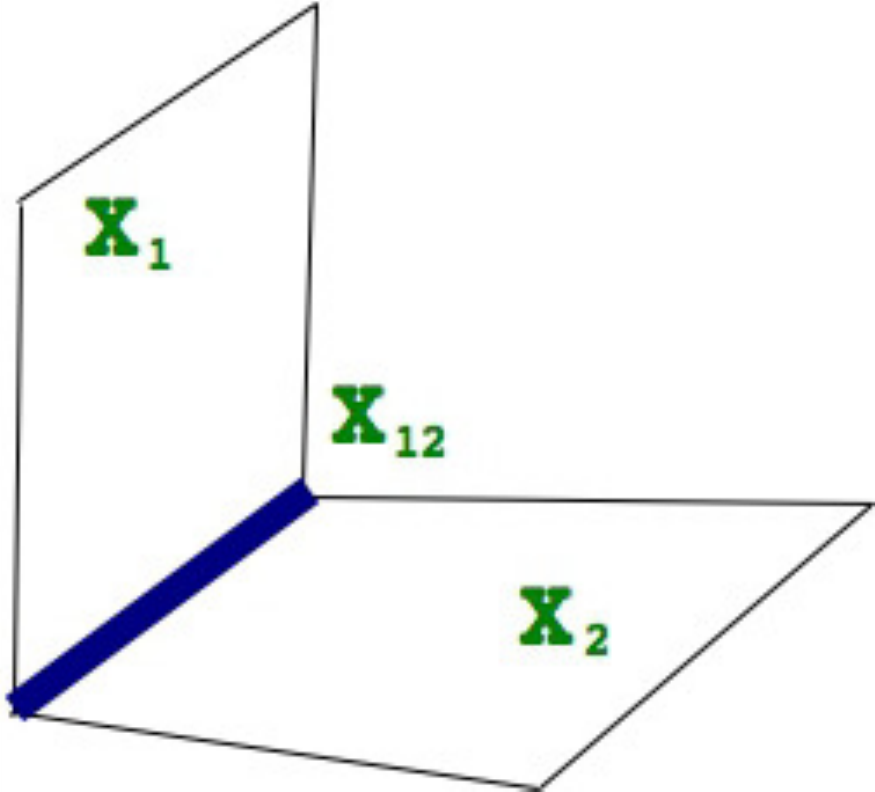}}

\bigskip

The simplest such example can be found in quantum mechanics, see appendix A.

While the maps (morphisms) between two quantum field theories come from the codimension one defects, the categorification suggests the studies of defects of all possible codimensions. In this paper we shall be mostly discussing the codimension two defects. 

The paper is organized as follows. The section $\bf 2$ recalls and clarifies some of the well-known facts about the two dimensional gauge theories and their point-like defects, as well as their three dimensional lifts and line defects. In particular, we discuss the moduli spaces of flat connections on punctured Riemann surfaces, and relate integration over these spaces to supersymmetric two dimensional Yang-Mills theory with point-like defects. The section $\bf 3$ adds matter to these theories
to make them gauged sigma models, then embeds these theories as a specific low-energy limit of compactified higher dimensional gauge theories. The sigma models on the moduli spaces of flat connections on punctured surfaces, in this context, become the four dimensional gauge theories with surface defects. The section $\bf 4$ introduces the supersymmetric partition functions, defined as a limit of twisted Witten index of a higher dimensional theory. We discuss equivariant index of Dirac operator on the moduli spaces of solutions to generalized Hitchin equations, quiver theories, and specific examples of  sigma models on Grassmanians and vector bundles over them. The section $\bf 5$ reviews a noncommutative gauge theory approach to the same problems. One views the gauge theory on a noncommutative space as a deformation of the original theory. The supersymmetric partition function has no continuous dependence on the noncommutativity parameters, allowing one to reduce the computation to a combinatorial formula. The section $\bf 6$ studies in detail three constructions of surface defects: the orbifold defects, the quiver defects, and the folded defects. The section $\bf 7$ discusses a specific class of local operators in two dimensional gauged linear sigma models, the $qq$-characters, which can also be viewed as a particular case of the codimension two defects. These operators are useful in revealing the hidden symmetries of quantum field theories we are studying. 
The concluding section $\bf 8$ discusses the topics we left out. 

\section{Examples in two dimensions}

Consider the two dimensional Yang-Mills theory with the gauge group $G$, a simple Lie group. In the cohomological field theory formulation is has the fields $A, {\psi}, {\sigma}$ with $A$ a connection, $\psi$ an adjoint-valued fermion one form,  and ${\sigma}$ and adjoint scalar, the electric field of the first order formalism. 
One assigns a $U(1)$-charge (ghost number) to these fields, with $A$ being neutral, $\psi$ of charge $1$ and $\sigma$ of charge $2$. 
The action is (we denote by $\Tr$ the Killing
form on ${\mathfrak g} = {\rm Lie}(G)$)
\beq
S_{YM} = - \frac{{\ii}t}{4\pi^2} \int_{\Sigma} {\Tr} ({\sigma}F_{A} + {\frac 12}{\psi} \wedge {\psi}) +  \frac{u}{8{\pi}^{2}} \int_{\Sigma}\, {\Tr}{\sigma}^{2}\, {\rm vol}_{\Sigma} \, 
\label{eq:2dym}
\eeq 
where
\beq
\frac{u}{t^2} = g^{2}_{\rm ym}
\eeq
is the Yang-Mills coupling constant.  
The theory \eqref{eq:2dym} has the fermionic symmetry, of ghost number $1$:
\beq
{\delta}A = {\psi}\, , \qquad {\delta}{\psi} = d_{A}{\sigma}\, , \qquad {\delta}{\sigma} = 0 
\label{eq:geqc}
\eeq 
representing the equivariant de Rham differential in the Cartan model of the equivariant cohomology of the space of gauge fields, with the symmetry group being the group of gauge transformations. The cohomology of $\delta$ in the space of local operators is isomorphic to the ring of $G$-invariant polynomials on the Lie algebra ${\mathfrak g} = {\rm Lie}(G)$, represented by the operators
\beq
{\CalO}_{P}^{(0)}(x) = P \left( {\sigma}(x) \right) 
\label{eq:0obs}
\eeq
The $\delta$-symmetry defines the so-called descent procedure, 
\beq
{\delta}{\CalO}^{(i)}_{P} = d{\CalO}^{(i-1)}_{P}
\eeq
which
associates to every local operator \eqref{eq:0obs} and a homology class
${\alpha} \in H_{i}({\Sigma}, {\BZ})$ the $\delta$-closed observables 
\beq
\int_{\alpha} {\CalO}^{(i)}_{P} \, , \qquad i = 0,1,2 \ ,
\label{eq:iobs}
\eeq
which we shall call the integrated $i$-observables. 
The action \eqref{eq:2dym} is actually a sum of a two-observable and a `smeared' zero-observable. The operators ${\Tr}{\sigma}^{n}$ have the ghost number $2n$,  the $i$-observable ${\CalO}^{(i)}_{{\Tr}{\sigma}^{n}}$ have the ghost number $2n-i$.

When $\Sigma = {\Sigma}_{g}$ is a genus $g$ compact Riemann surface, the partition
function of the theory \eqref{eq:2dym} , up to the nonperturbative in 
$g_{\rm ym}^{2}$ terms, would be equal to:
\beq
Z (t, u ) \sim  \int_{{\CalM}_{g}(G)} \, e^{t {\omega} + u {\Theta}} \ = \ \sum_{k=0}^{[d/2]}\  \frac{t^{d-2k} u^{k}}{k!(d-2k)!}\ \int_{{\CalM}_{g}(G)}  \  {\omega}^{d-2k} \wedge  {\Theta}^{k}\, , 
\eeq
the integral over the moduli space ${\CalM}_{g}(G)$ of flat $G$-connections on $\Sigma_{g}$, of a product of the symplectic form $\omega$ and a certain Pontryagin class $\Theta$ \cite{W}. 
The $U(1)$ ghost number symmetry has a gravitational anomaly, so that on the worldsheet $\Sigma$ only the operators of the total ghost number $2d \equiv - {\rm dim}(G) {\chi}({\Sigma})$ may have the non-zero expectation value.  
The ghost number anomaly coincides with the dimension of the moduli space of flat connections
\beq
{\rm dim}{\CalM}_{g}(G) = - {\chi}({\Sigma}_{g}) {\rm dim}(G) \, ,
\eeq
 where
 \beq
 {\CalM}_{g}(G) = {\rm Hom}({\pi}_{1}({\Sigma}_{g}) , G)/G
 \eeq
Unfortunately, the moduli space is not smooth (unless one turns on a discrete 't Hooft flux), so that the path integral actually localizes onto the
integral over the non-compact singular space of solutions to the equations
\beq
F_{A} = 0, \quad d_{A}{\sigma} = 0 
\eeq
making the $u$-dependence of $Z (t, u)$ non-analytic. 

We shall now couple the two-dimensional gauge theory
to a zero-dimensional one. 
The latter is a theory of integrals over the spaces
with $G$-symmetry. The building blocks of such spaces, in some sense, are
the coadjoint orbits. 
Let $T \subset G$ be the maximal torus, and ${\mathfrak t} = {\rm Lie}(T) \subset {\mathfrak g}$ the Cartan subalgebra. For ${\nu} \in {\mathfrak g}^{*}$ let $O_{\nu} \subset {\mathfrak g}^{*}$ denote the coadjoint orbit:
\beq
O_{\nu} = \{ \, Ad_{g}^{*}({\nu}) \, | \, g \in G \, \}
\eeq
Let ${\varpi}_{\nu} \in {\Omega}^{2}(O_{\nu})$ denote Kirillov-Kostant-Souriau form, the canonical symplectic form on $O_{\nu}$, and 
\beq
{\mu}_{\nu} : O_{\nu} \longrightarrow {\mathfrak g}^{*}
\eeq
the moment map, which coincides with the embedding map. Finally, let
\beq
{\CalV}_{\nu}({\sigma}) = \int_{O_{\nu}}  \, e^{{\varpi}_{\nu}+{\ii} {\mu}_{\nu} ( {\sigma} )}
\label{eq:eqvol}
\eeq
be the equivariant symplectic volume of $O_{\nu}$, an $Ad$-invariant function of $\sigma \in {\mathfrak g}$. One can compute ${\CalV}_{\nu}({\sigma})$ rather explicitly. Choose a representative of $\sigma$ in the Cartan subalgebra. Let us denote it by the same letter.  For the generic $\nu$, $O_{\nu} \approx G/T$, and generic $\sigma \in {\mathfrak t}$:
\beq
{\CalV}_{\nu}({\sigma}) = \frac{1}{\prod\limits_{{\alpha} > 0} {\alpha} ( {\sigma}  )}  \sum_{w \in W} (-1)^{w} e^{{\ii} {\nu} ({\sigma}^{w} )}
\label{eq:calv}
\eeq
where ${\alpha} \in {\mathfrak t}^{*}$ are the roots of $\mathfrak g$, $W$ is the Weyl group of $G$, and ${\sigma}^{w}$ is the
$w$-transformed $\sigma$. For example, for $G = SU(n)$, $T \approx U(1)^{n-1}$, $W = S(n)$, the symmetric group, ${\nu}$ and ${\sigma}$ are diagonal matrices, and the condition that $\nu$ is generic means that the eigenvalues ${\nu}_{i}, i = 1, \ldots , n$ of $\nu$ are distinct. If, in addition, $\sigma$ is generic, i.e. its eigenvalues ${\sigma}_{i}$ are distinct, then \eqref{eq:calv} specializes to:
\beq
{\CalV}_{{\nu}_{1}, \ldots, {\nu}_{n}}({\sigma}_{1}, \ldots , {\sigma}_{n}) = \frac{\ {\rm Det} \left\Vert e^{{\ii}{\sigma}_{i}{\nu}_{j}} \right\Vert_{i,j=1}^{n}}{\prod\limits_{i < j} ({\sigma}_{i} - {\sigma}_{j})}
\label{eq:suniz}
\eeq
the value of the integral, computed by A.~Kirillov, Harish-Chandra, C.~Itzykson and J.B.~Zuber and many others, in various contexts. The formula \eqref{eq:suniz} for non generic $\nu$, e.g. for $G = SU(n)$ when some of its eigenvalues coincide, gives zero. More precisely, the symplectic volume of $O_{\nu}$, the ${\sigma} \to 0$ limit of \eqref{eq:suniz} is equal to
\beq
v_{\nu} = \prod_{{\alpha} > 0} \, \langle {\alpha}, {\nu} \rangle
\eeq
and vanishes whenever ${\nu}$ falls onto a wall of a Weyl chamber.
However this zero is merely a reflection of the lower dimensionality of the corresponding orbit $O_{\nu} = G/G_{\nu}$. One can take a limit of the normalized equivariant volume ${\CalV}_{\nu}/v_{\nu}$ for $\nu$ approaching the wall of the chamber, and arrive at the 
fixed point formula given by the sum over the coset
$W/W_{\nu}$ with $W_{\nu}$ the Weyl group of the stabilizer $G_{\nu}$. For example, for $G = SU(n)$,  if the multiplicities of the eigenvalues of ${\nu}$ are equal to $(n_{1}, \ldots , n_{l})$, with $n_{1} + \ldots + n_{l} = n$, then $G_{\nu} = S\left( U(n_{1}) \times \ldots \times U(n_{l}) \right)$, $W_{\nu} = S(n_{1}) \times \ldots \times S(n_{l})$. The case $(k, n-k)$ corresponds to $O_{\nu} \approx Gr(k,n)$, the Grassmanian of $k$-planes in ${\BC}^{n}$, the $SU(n)$-orbit of a matrix of the form
\beq
{\nu} \cdot 1_{n} - I  I^{\dagger}  \, \qquad  I: {\BC}^{k} \to {\BC}^{n} \, , \qquad I^{\dagger} I =  \frac{n}{k} \, {\nu} \, \cdot 1_{k}\, , \qquad {\nu} > 0
\eeq
The corresponding integral 
\beq
{\CalV}_{\nu} ({\sigma}_{1}, \ldots , {\sigma}_{n}) = \sum\limits_{I \subset [n],\, {\#}I = k} \
\prod\limits_{i \in I} \, \frac{e^{{\ii}{\nu} {\sigma}_{i}}}{ \prod\limits_{j \notin I} ({\sigma}_{j} - {\sigma}_{i})} 
\label{eq:vzkn}
\eeq
The general case is not much more complicated: let ${\nu}_{a}$ be the eigenvalue of $\nu$ of multiplicity $n_{a}$, $a = 1, \ldots , l$, with
\beq
\sum_{a=1}^{l} n_{a}  = n \, , \qquad \sum_{a=1}^{l} n_{a} {\nu}_{a} = 0 \, , 
\eeq
then
\beq
{\CalV}_{\vec\nu} ({\sigma}_{1}, \ldots , {\sigma}_{n}) = \sum_{I_{1}\, \sqcup\, I_{2}\, \sqcup\, \ldots \, \sqcup\, I_{l} = [n]}\ \frac{\prod\limits_{a=1}^{l} \prod\limits_{i \in I_{a}} \, {\exp}\ {\ii} {\nu}_{a}  {\sigma}_{i}}{\prod\limits_{1\leq a < b \leq l} \prod\limits_{i \in I_{a}} \prod\limits_{j \in I_{b}} ( {\sigma}_{i}  - {\sigma}_{j} ) }
\eeq
with ${\#}I_{a} = n_{a}$. Note that the denominator has the ghost number 
\beq
2 \sum_{a<b} n_{a}n_{b} = {\rm dim}G/G_{\nu}
\label{eq:dimggnu}
\eeq
Now let ${\Sigma} = {\Sigma}_{g,n}$ be the genus $g$ Riemann surface with $n$ punctures $x_{1}, x_{2}, \ldots , x_{n} \in \Sigma$. Let us assign to the puncture $x_i$  a conjugacy class $[u_{i}] \in G/Ad(G)$ represented by $u_{i} = e^{\frac{2{\pi}{\ii}}{t} {\nu}_{i}} \in T$,  and consider the moduli space ${\CalM}_{g,n}(G; [u_{1}], \ldots , [u_{n}])$ of flat connections with the conjugacy class of holonomy around $x_{i}$ being $[u_{i}]$. 
For generic $[u_{i}]$, and for ${\chi}({\Sigma}_{g,n}) = 2 - 2g -n < 0$ this moduli space is a manifold of dimension
\beq
{\rm dim}{\CalM}_{g,n}(G; [u_{1}], \ldots , [u_{n}]) = 2{\rm dim}(G) (g-1) + \sum_{i=1}^{n} {\rm dim}O_{{\nu}_{i}} = - {\chi}({\Sigma}_{g,n}) {\rm dim}(G) - \sum_{i=1}^{n} {\rm dim}(G_{{\nu}_{i}} )
\eeq
{}The theory \eqref{eq:2dym} viewed perturbatively in the parameters $t, u$, can be embedded into the ${\CalN}=(2,2)$ supersymmetric Yang-Mills theory. The latter, with the $B$-type topological  twist, has additional fields: bosonic ${\bar\sigma}, H$, fermionic ${\eta}, {\chi}$, all scalars valued in the adjoint representation of $G$. In the physical theory, $\bar\sigma$ is complex conjugate of $\sigma$, in the mathematical applications on often takes $\sigma$ and $\bar\sigma$ to be real independent fields.  

{}Define the Wilson point operators, cf. \cite{Blau:1993hj}:
\beq
{\CalO}_{\nu}(x) = {\CalV}_{\nu} \left( {\sigma}(x) \right)
\eeq
On the one hand, these operators are simply a sophisticated linear combination of the more conventional ${\Tr} ( {\sigma}^{n} )$ operators. 

{}On the other hand, they generate a codimension two defect, 
\beq
F_{A} \sim  \sum_{i=1}^{n} J_{{\nu}_{i}}\, {\delta}_{x_{i}}^{(2)}
\label{eq:defop}
\eeq
where $J_{{\nu}_{i}} \in {\mathfrak g}$ belongs to the (co)adjoint orbit of ${\nu}_{i} \in {\mathfrak t}^{*}$.  
Indeed, add the ${\CalQ}$-exact term to the action of the ${\CalN}=(2,2)$ super-Yang-Mills theory of the form 
\beq
{\ii}{\bar t} {\CalQ} \int_{\Sigma} {\Tr} \left( {\bar\sigma} {\chi}  \right) 
\eeq
and take the limit ${\bar t} \to \infty$. In this limit the fields ${\bar\sigma}, H$
${\chi}$, $\eta$ decouple, and one is left with the `physical' Yang-Mills fields
$\sigma$, $A$, $\psi$ without kinetic term for the fermions $\psi$. 
Now, integrating out $\sigma$ in the limit of small $u$ one gets precisely
\eqref{eq:defop}. 

$\bullet$

To see that this is something one is really familiar with, let us deform the 
two dimensional 
 Yang-Mills theory to the gauged $G/G$ Wess-Zumino-Witten model:
\begin{multline}
S_{GWZW} = -\frac{k}{8{\pi}} \int_{\Sigma}\, g^{-1}d_{A} g \wedge \star g^{-1}d_{A}g  - {\ii} {\Gamma}(g, A) = \\
k \, S_{WZW}(g) - \frac{k}{2\pi} \int_{\Sigma}\, d^{2}z\, {\Tr}A_{z}{\pa}_{\zb}g g^{-1} + 
\frac{k}{2\pi} \int_{\Sigma}\, d^{2}z\, {\Tr}A_{\zb} \left( g^{-1}{\pa}_{z}g + g^{-1}A_{z} g - A_{z} \right) \\
\label{eq:gwzwac}
\end{multline}
where ${\Gamma}(g,A)$ is the gauge invariant extension of the Wess-Zumino term:
\beq
{\Gamma}(g) = \frac{1}{12{\pi}} \int_{{\pa}^{-1}{\Sigma}} {\Tr} \left( g^{-1} dg \right)^{3}
\eeq
Then the Wilson point operators deform to the operators (cf. \cite{BT, Ge, NPhD, BLN}):
\beq
{\CalO}_{\lambda}(x) = {\Tr}_{R_{\lambda}} \,  g(x)  
\label{eq:locopgwzw}
\eeq
By lifting the gauged WZW theory to the three dimensional 
Chern-Simons theory on $S^{1} \times {\Sigma}_g$ one can make the whole construction supersymmetric, with the topological supersymmetry reducing to that of the topological Yang-Mills theory in two dimensions \cite{NPhD, BLN}:
\beq
S_{CS} = \frac{k}{8{\pi}} \int_{S^{1} \times {\Sigma}} {\Tr} \left( A\wedge dA + {\frac 23} A\wedge A \wedge A \right)
\eeq
The operators \eqref{eq:locopgwzw} correspond to the `vertical' Wilson loops:
\beq
 \left\langle\, \prod_{i} {\CalO}_{{\lambda}_{i}} (x_{i}) \, \right\rangle_{GWZW, \Sigma} = \left\langle\ 
\prod_{i} {\Tr}_{R_{{\lambda}_{i}}} \, P{\rm exp} \oint_{S^{1}_{x_{i}}} \, A\ \right\rangle_{CS, S^{1} \times \Sigma}
\label{eq:csvw}
\eeq 
Now, the behavior \eqref{eq:defop} of the curvature  is the well-known response of the gauge field on the magnetic source. In Chern-Simons theory the electric sources (Wilson loops) are at the same time magnetic sources, as is well-known. 

The partition function computes for $k> 0$ the dimension of the space of conformal blocks,
\beq
Z_{g,n} (G;  [u_{1}], \ldots , [u_{n}]) = \sum_{i} \, (-1)^{i} \, {\rm dim} H^{i} \left({\CalM}_{g,n}(G; [u_{1}], \ldots , [u_{n}]) , {\CalL}^{\otimes k} \right) \ , 
\eeq
the celebrated Verlinde formula \cite{Ve, BT, Ge, W5}. 

The codimension two defects in Chern-Simons theory and in the gauged WZW theory correspond to the vertex operators in the two dimensional conformal field theory, which lives on the boundary of the three dimensional space. It was the dream of finding the analogues of such operators in the higher dimensional analogues of the two dimensional conformal field theories, such as WZW$_{4}$, that lead to the studies of surface operators in four dimensional supersymmetric gauge theories \cite{Losev:1995cr}\footnote{In order to count the four dimensional generalizations of WZW conformal blocks the supersymmetric localization of five dimensional gauge theory compactified on a circle has been developed \cite{NPhD}. Recently these concepts have resurfaced in \cite{Barns-Graham:2017zpv}}. The idea, presented in my lecture at the IAS program `Langlands Program and Physics' in the spring of 2004, was to couple the supersymmetric sigma model with the target space having a $G$-symmetry, to the four dimensional supersymmetric gauge theory with the gauge group $G$. 

\section{Between two and four dimensions}

In this section we pass from gauge theories in two dimensions to gauged 
sigma models, then generalize sigma models to four dimensional gauge theories. 

\subsection{Two dimensional gauged linear sigma models}

A two dimensional $\CalX$-sigma model on a Riemann surface $C$ is a quantum field theory whose fields are the maps ${\phi}: C \to {\CalX}$, to some fixed target space $\CalX$, with the Riemannian metric ${\rm G}$. The action, typically, contains a Dirichlet energy term, plus other terms, such as the $B$-field, a tachyon $\rm T$, and a dilaton $\Phi$ couplings:
\beq
S = \int_{C} \, {\rm vol}_C(h) \, \left( \, {\tr} \left( h^{-1} {\phi}^{*}{\rm G} \right) + {\phi}^{*}{\rm T} \right) + \int_{C} {\phi}^{*}B + \int_{C} {\CalR}_{C}(h) {\phi}^{*}{\Phi}
\label{eq:sigmac}
\eeq
Classically, the symmetries of $({\CalX}, {\rm G}, B, {\rm T}, \Phi)$ are the global symmetries of the sigma model. Gauging a subgroup of the group of symmetries produces another quantum field theory. Quantum mechanically, some of the classical symmetries may be anomalous, and cannot be consistently gauged. 

The gauged WZW model of the previous section is an example with the group manifold as the target space ${\CalX} = G$, with the left-right invariant metric ${\rm G} = {\Tr} (g^{-1}dg)^2$, the $B$-field related to the integral generator of $H^3 (G, {\BZ})$,  $dB =  
\frac{k}{12{\pi}} {\Tr} \left( g^{-1} dg \right)^{3}$, and the symmetry being the adjoint action of $G$ on itself. The purely left or right $G$-action on itself, albeit a symmetry of the metric $\rm G$, is not a symmetry of the $B$ field.

(Super)symmetry requirements limit the choices of ${\rm G}, B, {\Phi}, T$ etc. In particular ${\CalN}=(2,2)$ supersymmetry forces the target space to be generalized K\"ahler. Let us further restrict ourselves with the K\"ahler case. 
Let $(X^i)$ denote the local holomorphic coordinates on ${\CalX}$, and  
 $\omega = \frac{1}{2\ii} {\rm G}_{i{\bar j}} dX^{i} \wedge dX^{\bar j}$ the corresponding symplectic form. Suppose, in addition, that $({\CalX}, {\rm G}, {\omega})$ has the group $G$ of  symmetries (isometries of $\rm G$, symplectomorphisms of $\omega$), generated by the vector fields $V_{a}$, $a = 1, \ldots , {\rm dim}{\mathfrak g}$, with the moment map $\mu = ({\mu}^{a})$:
 \beq
 \iota_{V_{a}} {\omega} = d{\mu}^{a}\, , \qquad \{ {\mu}^{a}, {\mu}^{b} \} = f^{ab}_{c} {\mu}^{c} = {\omega} (V_{a}, V_{b})
 \eeq
 Gauging the $G$-symmetry in the supersymmetric fashion amounts to coupling the sigma model to the ${\CalN}=(2,2)$ supersymmetric gauge theory. The vector multiplet
 contains, in addition to the gauge field $A = (A^{a})$ a complex adjoint-valued scalar $\sigma$, $\bar\sigma$. 
 The theory, in the twisted form, has the nilpotent supercharge, which acts on the basic fields $(A, \psi, \sigma, \bar\sigma, \eta, \chi, D)$, $(X^{\mu}, {\rho}^{\mu}, {\chi}_{iz}, h_{iz}, {\chi}_{\bar i \bar z}, h_{\bar i \bar z})$ of the theory as follows:
 \beq
 \begin{aligned}
 & \qquad {\rm adjoint\ valued} \\
 & {\delta}A = {\psi}\, , \qquad {\delta}{\psi} = D_{A}{\sigma} \, , \qquad {\delta}{\sigma} = 0 \\
 & {\delta}{\bar\sigma} = {\eta}\, , \qquad {\delta}{\eta} = [ {\bar\sigma}, {\sigma} ] \\
 & {\delta}{\chi} = D\, , \qquad {\delta}D = [ {\chi}, {\sigma} ] \\
 & \qquad {\rm non-linear\ sigma\ model} \\
 & {\delta}X^{\mu} = {\rho}^{\mu}\, , \qquad {\delta}{\rho}^{\mu} = V^{\mu} ({\sigma}) = V^{\mu}_{a} {\sigma}^{a} \\
& {\delta}{\chi}_{iz} = h_{iz} \, , \qquad {\delta}h_{iz} = - {\pa}_{i} V^{j}({\sigma}) {\chi}_{jz} \\
& {\delta}{\chi}_{{\bar i}{\bar z}} = h_{{\bar i}{\bar z}} \, , \qquad {\delta}h_{{\bar i}{\bar z}} = - {\bar\pa}_{\bar i} V^{\bar j}({\sigma}) {\chi}_{{\bar j}{\bar z}} \\
\end{aligned}
\eeq 
The action of the theory is taken to be:
\beq
S = \int_{C} {\CalO}^{(2)}_{\omega+{\mu}} + {\delta} \int_{C} {\bf\Psi} 
\label{eq:sco2}
\eeq
where 
\beq
{\CalO}^{(2)}_{\omega+{\mu}} = \frac 12 {\omega}_{\mu\nu} {\nabla}_{A}X^{\mu} \wedge {\nabla}_{A}X^{\nu} + {\ii} {\mu} (F_{A})
\label{eq:om2obs}
\eeq
is the $2$-observable, corresponding to the $0$-observable 
\beq
{\CalO}^{(0)}_{\omega+{\mu}} = \frac 12 {\omega}_{\mu\nu}  {\rho}^{\mu} {\rho}^{\nu} + {\ii} {\mu}({\sigma}) 
\label{eq:omobs}
\eeq
which are included in the tower $d{\CalO}^{(i)} = {\delta}{\CalO}^{(i+1)}$,
\beq
{\CalO}^{(1)}_{\omega+{\mu}} = {\omega}_{\mu\nu}  {\rho}^{\mu} {\nabla}_{A}X^{\nu} + 
{\mu}({\psi})
\label{eq:om1obs}
\eeq
In the formulae above we used the notation
\beq
{\nabla}_{A}X^{\mu} = dX^{\mu} + V^{\mu}(A) \ .
\eeq 
 One can concisely rewrite \eqref{eq:omobs}, \eqref{eq:om1obs}, \eqref{eq:om2obs} as
 \beq
 {\CalO}^{(0)}_{\omega+{\mu}} +   {\CalO}^{(1)}_{\omega+{\mu}} +{\CalO}^{(2)}_{\omega+{\mu}} 
 = \frac 12 {\omega}_{\mu\nu} \left( {\rho}^{\mu} + {\nabla}_{A}X^{\mu} \right) \left( {\rho}^{\nu} + {\nabla}_{A}X^{\nu} \right)  + {\mu} \left( {\sigma} + {\psi} + F_{A} \right)
 \eeq
 More generally, let
 \beq
 f(X, {\rho}, {\sigma})
 \eeq
 be the equivariantly closed differential form on ${\CalX}$, then
 \beq
 {\CalO}^{(0)}_{f} +   {\CalO}^{(1)}_{f} +{\CalO}^{(2)}_{f}  = f ( X, {\rho} + {\nabla}_{A}X, {\sigma} + {\psi} + F_{A} ) 
 \label{eq:eqfrm}
 \eeq
is ${\delta}$-closed observable in the sigma model. 
  
{}The bosonic part of the action of the coupled theory has the form:
 \begin{multline}
 \int_{C} \, {\rm G}_{i{\bar j}} \left( dX^i + V^{i}_{a} A^{a} \right) \wedge \star \left( dX^{\bar j} + V^{\bar j}_{a} A^{a} \right) \, + \,  
 {\Tr} {\mu}^{2} \, + \, {\rm G}_{i{\bar j}} V^{i}_{a}V^{\bar j}_{b} {\sigma}^{a}{\bar\sigma}^{b} + \\
 {\Tr} D_{A}{\sigma} \wedge \star D_{A}{\bar \sigma} + {\Tr} [{\sigma}, {\bar\sigma}]^{2} + {\Tr} F_{A} \wedge \star F_{A}
 \end{multline}
The supersymmetric localization locus (it corresponds to the $B$-model twist of the $\sigma, \bar\sigma$-multiplet, and the $A$-model twist of the $X^{i}, X^{\bar j}$-multiplet) reads:
\beq
\begin{aligned}
& \star F_{A} + {\mu} = 0 \\
& {\bar\pa}_{\zb} X^{i} + A^{a}_{\zb} V^{i}_{a} = 0 \\ 
& {\sigma}^{a} V^{i}_{a} = 0 \\
& D_{A}{\sigma} = 0 \\
& [ {\sigma}, {\bar\sigma} ] = 0 \\
\end{aligned}
\eeq

\subsection{From sigma models to gauge theories}

An interesting class of sigma models in two dimensions have the moduli spaces of solutions to some field equations as their target spaces. Such models arise as limits of higher dimensional field theories. When $X = {\CalM}_{g}(G)$ is the moduli space of flat $G$-connections  on a genus $g$ Riemann surface $\Sigma_{g}$, the theory \eqref{eq:sigmac} is an approximation to the four dimensional gauge theory with gauge group $G$, with the Euclidean spacetime $M^4 = {\Sigma}_{g} \times C$, in the limit, where the size of $\Sigma_{g}$ is much smaller then that of $C$.  The metric ${\rm G}$ on ${\CalM}_{g}(G)$ is determined by the four dimensional gauge coupling ${\rm g}_{\rm ym}^{2}$, the $B$-field is induced from the four dimensional theta angle $\vartheta$. In the nonsupersymmetric theory, quantum corrections induce the tachyon, which corresponds to a potential on the moduli space, lifting the moduli.

However, if one starts with the supersymmetric gauge theory in four dimensions  \cite{BJSV}, e.g. ${\CalN}=2$ super-Yang-Mills theory, twisted or partially twisted along $\Sigma_{g}$, one gets a twisted or simply supersymmetric sigma model in two dimensions, with $C$ as a worldsheet (of course, for curved $C$ the supersymmetry is broken unless one gives appropriate expectation values to some of the supergravity fields, e.g. the $R$-symmetry gauge field, which amounts to the traditional twisting). 

Specifically, the pure ${\CalN}=2$ gauge theory in four dimensions with the canonical (Donaldson-Witten) twist gives rise  \cite{BJSV} to the $A$ model in two dimensions, whose target space is the moduli space ${\CalM}_{g}(G)$ of flat connections. The analysis of  \cite{BJSV} breaks down when the moduli space is singular, which means one has to turn on a 't Hooft flux. 

However\footnote{Author's remark during C.~Vafa lecture at NHETC at Rutgers presenting  \cite{BJSV} in 1995}, one can perfectly well study the $A$ model with the moduli space of flat connections  on the punctured Riemann surface. These moduli spaces are `nicer' then those without punctures, and even in genus $g=0$ one can get a smooth space or at least a space with at most orbifold singularities. What kind of four dimensional gauge theory would lead to such a model?

\subsection{The surface defects}

To answer this question we need to learn a little bit about the complex structure of
\[ {\CalM}_{g,n}(G ; [u_{1}], \ldots , [u_{n}]) \ . \]
Of course, the $A$ model does not care about the choice of the complex structure, but the localization locus of the path integral does. By looking at the localization locus we shall be able to reconstruct the localization locus of the four dimensional theory on $C \times {\Sigma}_{g}$, which gives the sigma model in the limit of vanishing area of $\Sigma_{g}$. So we pick a complex structure on $\Sigma_{g}$, with local coordinates $z, \zb$. Then the space of all gauge fields ${\CalA}_{\Sigma_{g}}$ becomes an infinite dimensional affine space, with the coordinates $A_{\zb}(z, \zb)$. The complexification ${\CalG}_{\BC}$ of the gauge group acts on ${\CalA}_{\Sigma_{g}}$ holomorphically: 
\beq
A_{\zb} \sim g^{-1}A_{\zb} g + g^{-1} {\pa}_{\zb} g
\label{eq:azbg}
\eeq
The quotient ${\CalA}_{\Sigma_{g}}/{\CalG}_{\BC}$ is the moduli {\it stack} $Bun_{G_{\BC}}({\Sigma}_{g})$ of holomorphic $G_{\BC}$-bundles on ${\Sigma}_{g}$. However, there is a subset $Bun_{G_{\BC}}({\Sigma}_{g})^{\rm ss}$ of so-called {\it semi-stable} bundles, which is isomorphic to the moduli space ${\CalM}_{g}(G)$ of flat $G$-connections. Specifically, it means that ${\pa}_{\zb} + A_{\zb}$ represents the semi-stable $G_{\BC}$-bundle iff one can find $g$ in \eqref{eq:azbg} such that
\beq
[ g^{\dagger} \left( {\pa}_{z} + A_{z} \right)  (g^{\dagger})^{-1},\ g^{-1} \left( {\pa}_{\zb} + A_{\zb} \right)  g]  = 0
\label{eq:faaz}
\eeq
This is the content of Narasimhan-Seshadri's theorem. 

Now, with $n$ marked points $x_{1}, \ldots , x_{n} \in {\Sigma}_{g}$ one can define an additional structure: namely, one choose the so-called parabolic structure at each marked point, i.e. the reduction of the structure group $G_{\BC}$ to some parabolic subgroup $P_{i}$, $i= 1, \ldots , n$ (the subgroups may differ at different points). This modifies the definition of stability.  The analogue of the Narasimhan-Seshadri's theorem in this case identifies the moduli space $Bun_{G_{\BC}}\left({\Sigma}_{g}; (x_{i}, P_{i})_{i=1}^{n}\right)$ of semi-stable parabolic bundles with that of the moduli space of $G$-flat connections on the punctured surface $\Sigma_{g}$, with the fixed conjugacy classes $[u_{i}]$, $i = 1,\ldots , n$, of holonomy around the punctures. 
 
The parabolic subgroups $P_{i} \subset G_{\BC}$ are such that
\beq
O_{{\nu}_{i}} \approx G_{\BC}/P_{i}
\label{eq:complorb}
\eeq
as complex manifolds. 

Now, on an open dense subset of $Bun_{G_{\BC}}({\Sigma}_{g})$ the choice of the parabolic structure at the marked points is simply a choice of a point in $G_{\BC}/P_{i}$ for each $i = 1, \ldots , n$, so we get a forgetting map:
\beq
f_{n}: Bun_{G_{\BC}}\left({\Sigma}_{g}; (x_{i}, P_{i})_{i=1}^{n}\right) \longrightarrow Bun_{G_{\BC}}\left({\Sigma}_{g}\right)
\label{eq:forgetbung}
\eeq
associating with the parabolic bundle the underlying holomorphic bundle, and $n$ evaluation maps
\beq
e_{i}: Bun_{G_{\BC}}\left({\Sigma}_{g}; (x_{i}, P_{i})_{i=1}^{n}\right) \longrightarrow G_{\BC}/P_{i}
\label{eq:evi}
\eeq
sending the parabolic bundle to the corresponding generalized flag variety. 
Now, imagine taking the line bundles ${\CalL}_{i, \alpha}$ over $G_{\BC}/P_{i}$ generating the $K$-theory of the corresponding generalized flag variety, pull back their Chern classes using $e_{i}'s$ and push them forward using $f_{n}$. We'd get a cohomology class of $Bun_{G_{\BC}}\left({\Sigma}_{g}\right)$ which one may want to express using the observables of the two-dimensional Yang-Mills theory. The claim is that this is the product of the Wilson point operators we discussed before:
\beq
(f_{n})_{*} \bigwedge\limits_{i=1}^{n} e_{i}^{*} \left( e^{\sum\limits_{\alpha}\, {\nu}_{i, \alpha} c_{1}({\CalL}_{i,\alpha})} \right) \ \sim \ \prod\limits_{i=1}^{n} \, {\CalO}_{{\nu}_{i}} 
\label{eq:conui}
\eeq
Now we introduce $C$ and make everything vary holomorphically over $C$. 
A collection of holomorphic bundles over $\Sigma_{g}$, which is holomorphically parametrized by $C$ defines a holomorphic bundle over the complex surface $S = C \times {\Sigma}_{g}$. Moreover, if for each point $w \in C$ the corresponding bundle over $\Sigma_g$ is stable, then so is the bundle over $S$. The converse is not necessarily true. 
Stable bundles over $S$ correspond to the $G$-connections with anti-self-dual curvature:
\beq
F_{A}^{+} = \frac 12 \left( F_{A} + \star F_{A} \right) = 0
\label{eq:asd}
\eeq
The Hodge star $\star$ on two-forms knows about the conformal structure of $S$. The equations \eqref{eq:asd} mean that
\beq
F_{\zb \wb} = 0 \leftrightarrow {\pa}_{\wb} A_{\zb} = D_{\zb} A_{\wb}
\label{eq:02eq}
\eeq
i.e. the holomorphic bundle on $\Sigma_{g}$ defined by $D_{\zb} \equiv {\pa}_{\zb} + A_{\zb}$ varies holomorphically in $w$, up to isomorphisms. 
The equations \eqref{eq:asd} also imply:
\beq
h^{z\zb} F_{z\zb} + h^{w\wb} F_{w\wb} = 0 
\label{eq:11eq}
\eeq
where we assume the hermitian metric on $S$ of the split form: 
\beq
h_{z\zb} dzd{\zb} + h_{w\wb} dwd{\wb}
\eeq
Taking the limit of small $\Sigma_{g}$ means taking $h_{z\zb} \to 0$
while keeping $h_{w\wb}$ finite. In this limit the Eq. \eqref{eq:11eq} means that {\it for almost all} $w$ the curvature $F_{z\zb}$ vanishes. However, in a small region of the $w$ space the curvature $F_{z\zb}$ may remain finite. This is the {\it freckled instanton} phenomenon \cite{Losev:1999nt, Losev:1999tu}, well-known in the study of vortex equations. In fact, the four dimensional instanton equations \eqref{eq:11eq} are nothing but the BPS vortex equations written for the gauged linear (actually, affine) sigma model with worldsheet $C$ on $X = {\CalA}_{\Sigma_{g}}$ with the infinite-dimensional gauge group ${\CalG}$ of two dimensional gauge transformations. The equation $F^{0,2} = 0$ \eqref{eq:02eq} is the $A$-model localization locus in the presence of gauge symmetry, while the equation \eqref{eq:11eq} is the $D$-term moment map equation. 

Now let us add the parabolic structures. The fact that they also vary holomorphically in $w$ means that we have a collection of holomorphic maps 
\beq
{\varphi}_{i} : C \longrightarrow  G_{\BC}/P_{i} \, , \qquad i = 1, \ldots , n
\eeq
So the four dimensional lift of the $A$ model on ${\CalM}_{g,n}(G ; [u_{1}], \ldots , [u_{n}])$
looks like a four dimensional ${\CalN}=2$ super-Yang-Mills theory coupled to the two dimensional $A$ model on 
\beq
X = \varprod\limits_{i=1}^{n}\,  G_{\BC}/P_{i}
\eeq
We can view the combined theory again as the gauged linear sigma model with
\beq
X = {\CalA}_{\Sigma_{g}} \ \times \ \varprod\limits_{i=1}^{n}\,  G_{\BC}/P_{i}
\eeq
where the gauged group is the infinite dimensional group ${\CalG}$ of gauge transformations on $\Sigma_{g}$. Then the $A$-model localization locus would amount to:
\beq
\begin{aligned} 
& F^{0,2} = 0 \qquad {\rm outside} \qquad C \times \{ \, x_{1}, \ldots x_{n} \, \} \\
& {\pa}_{\wb} {\xi}_{i} = 0 \, ,\qquad {\xi}_{i} = \ {\rm holomorphic\ coordinates\ on}\qquad  G_{\BC}/P_{i} \\
\end{aligned}
\label{eq:02eqp}
\eeq
and the $D$-term equation
\beq
h^{z\zb} \left( F_{z\zb} + \sum_{i=1}^{n} J_{{\nu}_{i}} {\delta}^{(2)}(x_{i}) \right) + h^{w\wb} F_{w\wb} = 0
\eeq
We have arrived at the surface operators in gauge theory. They were introduced in \cite{KM1, KM2}. See \cite{Gukov:2007ck,Gukov:2008sn} for the recent discussion in the context of supersymmetric gauge theories.

\section{Supersymmetric count and partition functions}

In this section we study two dimensional gauged linear sigma model with 
${\CalN}=(4,4)$ supersymmetry, which is softly broken to ${\CalN}=(2,2)$ by the
twisted mass corresponding to the specific $U(1)$-symmetry.

\subsection{Twisted Witten index and generalized Hitchin equations}

We want to compute the partition function of this theory in the $\Omega$-background
with the parameter ${\ve}$. In other words, we deform the theory
by first lifting it to three dimensions and then compactifying on an ${\BR}^{2}$ (or $D^{2}$)
-bundle over $S^1$ of small circumference $r$ with the rotation of the fiber by the angle ${\ve}r$. Then we take the $r \to 0$ limit to get back the two dimensional theory. In addition to the
${\BR}^{2}$-rotation, we twist the boundary conditions by the global gauge  and flavor 
\beq
\begin{aligned}
& {\rm g}_{\rm gauge} = e^{{\ii} r \, {\sigma}} \, , \qquad {\sigma} \in {\mathfrak t} \subset {\mathfrak g} \\
& {\rm g}_{\rm flavor} = e^{{\ii} r \, m} \, , \qquad m \in {\rm Lie}(G_{\rm flavor}) \\
\end{aligned}
\eeq
transformations. We compute
\beq
{\CalZ}_{3d} = {\Tr}\, \left( \, (-1)^{F}\, e^{-r{\hat H}} \, {\rm g}_{\rm gauge}\, {\rm g}_{\rm flavor} \, e^{{\ii}{\ve}r {\hat J}_{\rm rot}} \, \right)
\label{eq:3dpf}
\eeq
Here ${\hat J}_{\rm rot}$ is a combination of the generator of the ${\BR}^{2}$ rotation and the $R$-symmetry which makes the twisted boundary conditions compatible with some supersymmetry. The specific twist we employ makes the bosonic fields of vector multiplet into
$(A_{z} dz + A_{\zb}d{\zb}, {\Phi}_{z}dz + {\Phi}_{\zb}d{\zb}, 
{\sigma}, {\bar\sigma})$, while the bosonic fields of the hypermultplet  $(Q, {\tilde Q})$ and their complex conjugates $(Q^{*}, {\tilde Q}^{*})$ remain scalars. Here $A_{\zb}$ is the $(0,1)$-component of a $G$-gauge field, with the compact Lie group $G$, ie. a connection on a principal $G$-bundle ${\CalP} \to \Sigma$,  $\Phi_{z}dz$ is the $(1,0)$-form valued in the adjoint vector bundle,
\beq
{\Phi}_{z}dz \in {\Gamma} \left( {\Omega}^{1,0}_{\Sigma} \otimes {\mathfrak{g}}_{\CalP} \right) \, , 
\eeq 
$Q$ is a section of a vector bundle $R_{\CalP}$, associated with the (complex) representation $R$ of $G$, and ${\tilde Q}$ is a section of a vector bundle $R^{*}_{\CalP}$, associated with the dual representation. Let 
\beq
{\mu}_{\BR} \oplus {\mu}_{\BC} : R \oplus R^{*} \longrightarrow {\mathfrak g}^{*} \otimes \left( {\BR} \oplus {\BC} \right)
\label{eq:hypmom}
\eeq
be the hyperk\"ahler moment map
\beq
\begin{aligned}
& {\mu}_{\BR}^{\ba} = \left( T_{R}^{\ba} \right)_{i}^{j} \left( Q^{i} Q^{*}_{j} - {\tilde Q}_{j} {\tilde Q}^{*, i} \right) - {\zeta}_{\BR}^{\ba}
\\
& {\mu}_{\BC}^{\ba} = \left( T_{R}^{\ba} \right)_{i}^{j} Q^{i}  {\tilde Q}_{j}  - \zeta_{\BC}^{\ba}\\
\end{aligned}
\label{eq:hypmom2}
\eeq
where ${\zeta}_{\BR}, {\zeta}_{\BC}$ are the real and the complex Fayet-Iliopoulos terms,
valued in the centralizer of $\mathfrak g$. 
 
The partition function localizes onto the integral over  the moduli space ${\CalM}$ of solutions to the generalized vortex equations (recall that $Q^{\dagger} = Q^{*}, \ {\tilde Q}^{\dagger} = {\tilde Q}^{*}, {\Phi}_{z}^{\dagger} = {\Phi}_{\zb}$):
\beq
\begin{aligned}
& D_{\zb} Q \equiv {\partial}_{\zb} Q  + T_{R} (A_{\zb}) \cdot Q = \left( T_{R^{*}} ({\Phi}_{z}) \cdot {\tilde Q} \right)^{\dagger} \\
& D_{\zb} {\tilde Q} = \left( T_{R} ({\Phi}_{z}) \cdot {Q} \right)^{\dagger} \\
& D_{\zb} {\Phi}_{z}  +  {\mu}_{\BC}^{\dagger} {\rm vol}_{\Sigma} = 0 \\
& F_{z\zb} + [ {\Phi}_{z}, {\Phi}_{\zb} ] +  {\mu}_{\BR} {\rm vol}_{\Sigma} = 0 \\
\end{aligned}
\label{eq:genhit}
\eeq
modulo gauge transformations 
\beq
(A, {\Phi}, Q, {\tilde Q}) \mapsto (g^{-1}Ag + g^{-1}dg, g^{-1}{\Phi}g, T_{R}(g^{-1})\cdot Q, T_{R^{*}}(g^{-1})\cdot {\tilde Q}) \ .
\label{eq:gage}
\eeq 
More specifically, we wish to study the equivariant index of Dirac operator on ${\CalM}^{\rm framed}$, possibly twisted with a vector bundle. Here the superscript ``framed'' means that the gauge symmetry group is reduced, by requiring $g$ in \eqref{eq:gage} to tend to $1$ at infinity. Instead of solving the Eqs. \eqref{eq:genhit} and studying the Dirac operator on the space of solutions, we modify the Dirac operator on the ambient space $M$, by coupling it to the exterior powers ${\Lambda}^{i}E$ of the vector bundle  $E$ where the Eqs. \eqref{eq:genhit} take values: ${\dir} \longrightarrow {\dir} \otimes 1_{{\Lambda}^{i}E} + 1 \otimes {\Lambda}^{i}s$. Here we represent \eqref{eq:genhit} collectively by $s = 0$, with $s: M \to E$.  The indices of these operators are then summed over all values of $i$ with the sign $(-1)^{i}$. The result is the index of Dirac operator restricted onto the zero locus of $s$, since the Chern character of $\sum_{i} (-1)^{i} {\Lambda}^{i}E$ is the Poincare dual of the $s = 0$ locus. One can make a more detailed argument using the supersymmetric quantum mechanics representation.

\subsection{Quiver gauge theories}

In this paper the gauge group and the matter representation 
corresponds to a quiver $\gamma$. The gauge group
\beq
G_{\rm gauge} = \varprod_{i \in \Vg} U({\bN}_{i}), 
\label{eq:ggr}
\eeq
is the product of unitary groups corresponding to the vertices of the quiver, the representation:
\beq
R = \bigoplus\limits_{i \in \Vg} \, {\rm Hom}({\bM}_{i}, {\bN}_{i}) \, \oplus \, \ \bigoplus\limits_{e \in \Eg} \, {\rm Hom}({\bN}_{s(e)}, {\bN}_{t(e)}) 
\label{eq:hyper}
\eeq
with ${\bM}_i$, $i \in \Vg$, the spaces of multiplicities of the fundamental hypermultiplets. 

The computation of \eqref{eq:3dpf} can be done simply by enumerating all holomorphic functions of
$A_{\zb}(z, {\zb}), {\Phi}_{z}(z, {\zb}), Q(z,{\zb}), {\tilde Q}(z, {\zb})$ modulo linearized gauge symmetries and equations 
\eqref{eq:genhit}, treating $z, \zb$-dependence formally. This is a free field character, given by the plethystic exponent 
\beq
{\CalZ}_{3d} = {\exp} \, - \, \sum_{l=1}^{\infty} \, \frac{1}{l} 
F (q^{l}, {\rm g}_{\rm gauge}^{l}, {\rm g}_{\rm flavor}^{l}) 
\label{eq:plethexp}
\eeq
where\footnote{The denominator $(1-q)(1-q^{-1})$ comes from the $z, \zb$-dependence of the fields $A_{\zb}(z, {\zb})$, etc. e.g. the mode $z^{i}{\zb}^{j} d{\zb}$ contributes $q^{i}(q^{-1})^{j} q^{-1}$}
\beq
F (q, {\rm g}_{\rm gauge}, {\rm g}_{\rm flavor}) = 
\frac{{\chi}_{\rm fields} - {\chi}_{\rm symmetries} - {\chi}_{\rm equations}}{(1-q)(1-q^{-1})}
\eeq
with
\beq
\begin{aligned}
& {\chi}_{\rm fields} =  \sum_{i \in \Vg} \left\{  \left(  q^{-1} + q t^{2} \right) N_{i}N_{i}^{*} + t^{-1}
\left( M_{i}N_{i}^{*} + N_{i} M_{i}^{*} \right)  \right\} +
\sum_{e \in \Eg} \, t^{-1}\, \left( N_{s(e)}N_{t(e)}^{*} + N_{t(e)} N_{s(e)}^{*} \right)   \\
& {\chi}_{\rm symmetries} =  \sum_{i \in \Vg} \, N_{i}N_{i}^{*} \\
& {\chi}_{\rm equations} =  \sum_{i \in \Vg} \left\{ \, t^{2}\, N_{i}N_{i}^{*} \, + \, q^{-1} t^{-1}\, 
\left( M_{i}N_{i}^{*} + N_{i} M_{i}^{*} \right)   \right\} +
\sum_{e \in \Eg}\, q^{-1} t^{-1} \, \left( N_{s(e)}N_{t(e)}^{*} + N_{t(e)} N_{s(e)}^{*} \right)   \\
\end{aligned}
\eeq
Here $q = e^{{\ii}{\ve}r}$ is the rotation eigenvalue, $t = e^{{\ii}r u}$ is
the soft ${\CalN} = 4 \longrightarrow {\CalN}=2$ supersymmetry breaking  phase corresponding to the $U(1)$ symmetry under which both $Q$ and ${\tilde Q}$ have charges $+1$, while ${\Phi}_{z}$ has charge $-2$, 
\beq
\begin{aligned}
& N_{i} = {\Tr}_{{\bN}_{i}} \, {\rm g}_{\rm gauge}\, , \qquad
N_{i}^{*} = {\Tr}_{{\bN}_{i}} \, {\rm g}_{\rm gauge}^{\dagger}
\\
& M_{i} = {\Tr}_{{\bM}_{i}} \, {\rm g}_{\rm flavor}\, , \qquad
M_{i}^{*} = {\Tr}_{{\bM}_{i}} \, {\rm g}_{\rm flavor}^{\dagger} \\
\end{aligned}
\eeq
Recall the formula for local contribution to the equivariant index of Dirac operator:
\beq
\prod_{\alpha} \, \frac{1}{2\, {\rm sinh} \left( \frac{{\xi}_{\alpha}}{2} \right)} = e^{-\frac 12 \sum_{\alpha} \xi_{\alpha}}
\ {\exp} \ \left( - \, \sum_{l=1}^{\infty} \frac{1}{l} \sum_{\alpha} \, e^{-l {\xi}_{\alpha}} \right)
\label{eq:locind}
\eeq
Since changing the sign of $\xi_\alpha$ simply changes the overall sign of the contribution \eqref{eq:locind} we can, modulo the overall sign ambiguity, rewrite \eqref{eq:3dpf} in a compact form (by replacing some of the characters by their duals):
\beq
F \sim \frac{1-q t^{2}}{1-q} \left\{ \sum_{i\in \Vg} (t^{-1} M_{i}- N_{i})N_{i}^{*} + \sum_{e\in \Eg} t^{-1}N_{s(e)}N_{t(e)}^{*} \right\}
\eeq

In what follows we shall use the notation (cf. \cite{N5})
\beq
{\epsilon} \left[ F^{({\tau})} \right] = {\exp}\, \frac{d}{ds} \Biggr\vert_{s=0} \ \frac{1}{{\Gamma}(s)} \int_{0}^{\infty} \frac{d\tau}{\tau} \, {\tau}^{s}\, F^{({\tau})}
\eeq

In the limit $r \to 0$, the plethystic sum in \eqref{eq:3dpf} over $l$ becomes a proper time integral:
\beq
{\CalZ}_{2d}^{\rm pert} = {\epsilon} \left[  \frac{1- e^{{\tau}({\ve} + 2u)}}{1-e^{{\tau}{\ve}}} \, \left\{ \sum_{i\in \Vg} (e^{-{\tau}u} M_{i}^{({\tau})}- N_{i}^{({\tau})})N_{i}^{(-{\tau})} + \sum_{e\in \Eg} e^{-{\tau}u} N_{s(e)}^{({\tau})}N_{t(e)}^{(-{\tau})} \right\} \right]
\label{eq:pertpf}
\eeq
with 
\beq
N_{i}^{({\tau})} = \sum_{{\alpha} = 1}^{n_{i} } e^{{\tau} a_{i, \alpha}} \, , \qquad M_{i}^{({\tau})} = \sum_{f = 1}^{{\rm rk}{\bM}_{i}} e^{{\tau} m_{i, f}}
\eeq
(with $n_i = {\rm rk}{\bN}_{i}$)
which becomes a product of $\Gamma$-functions (cf. \cite{N5})
In \eqref{eq:pertpf} we put the superscript {\tiny pert} to stress the fact that this is only a perturbative, i.e. one-loop exact, contribution to the partition function. 

The full partition function is not much more complicated. Namely, we sum over the flux sectors, where the gauge field for each of the $U(1)$ subgroups of each of the $U({\bN}_{i})$ groups may have a non-vanishing flux $F_{i, \alpha}$, with 
\beq
\int_{{\BR}^{2}} F_{i,\alpha} = 2{\pi}{\ii} d_{i, \alpha} \, , \qquad d_{i, \alpha} \in {\BZ}
\eeq
The instanton contribution to the partition function is computed, e.g. with the help of the family index theorem, as:
\beq
{\CalZ}_{2d}  = \sum\limits_{\left( {\bf d}_{i} \right)_{i \in \Vg}} \ \left( \prod_{i \in \Vg} {\qe}_{i}^{|{\bf d}_{i}|} \right) \ \cdot \ {\epsilon} \left[  \frac{1- e^{{\tau}({\ve} + 2u)}}{1-e^{{\tau}{\ve}}} \, \left\{ \sum_{i\in \Vg} (e^{-{\tau}u} M_{i}^{({\tau})}- {\Sigma}_{i}^{({\tau})}){\Sigma}_{i}^{(-{\tau})} + \sum_{e\in \Eg} e^{-{\tau}u} {\Sigma}_{s(e)}^{({\tau})}{\Sigma}_{t(e)}^{(-{\tau})} \right\} \right]
\label{eq:instpf}
\eeq
with ${\bf d}_{i} = (d_{i, \alpha})_{\alpha \in [n_{i}]}$, ${\qe}_{i}$ being the exponentiated K\"ahler moduli, 
\beq
| {\bf d}_{i} | = \sum_{\alpha} d_{i, \alpha}
\eeq
and
\beq
\Sigma_{i}^{({\tau})} = \sum_{\alpha} e^{{\tau}{\sigma}_{i, \alpha}} 
\eeq
with 
\beq
{\sigma}_{i, {\alpha}} = a_{i, \alpha} + {\ve} d_{i, \alpha} \ .
\eeq
The range of the fluxes $d_{i, \alpha}$ actually depends on the 
stability data. In our examples below these will be simply $d_{i, \alpha} \geq 0$. 

By playing with the $u$ and $m_{i,f}$ parameters one can get the more general ${\CalN}=(2,2)$ quiver gauge theory partition function. In this case the quiver edges (now one allows multiple edges connecting two vertices, possibly with the opposite orientations) correspond to the bi-fundamental chiral multiplets, while the nodes come with two types of multiplicities, ${\bM}_{i, \pm}$, for the fundamental, and the anti-fundamental chiral multiplets, respectively. The partition function is equal to, then:
\begin{multline}
{\CalZ}_{2d}^{{\CalN} = 2}  =\\
 \sum\limits_{\left( {\bf d}_{i} \right)_{i \in \Vg}} \ \left( \prod_{i \in \Vg} {\qe}_{i}^{|{\bf d}_{i}|} \right) \ \cdot \ {\epsilon} \left[  \frac{1}{1-e^{{\tau}{\ve}}} \, \left\{ \sum_{i\in \Vg} \left( M_{i,+}^{({\tau})}{\Sigma}_{i}^{(-{\tau})} +
M_{i,-}^{(-{\tau})}{\Sigma}_{i}^{({\tau})} 
- {\Sigma}_{i}^{({\tau})}{\Sigma}_{i}^{(-{\tau})} \right) + \sum_{e\in \Eg} \, e^{{\tau}m_{e}} \, {\Sigma}_{s(e)}^{({\tau})}{\Sigma}_{t(e)}^{(-{\tau})} \right\} \right] \\
\cdot \, {\epsilon} \left[  \frac{1}{1-e^{{\tau}{\ve}}} \, \left\{ \sum_{i\in {\Vg}} \ e^{{\tau}{\ve}} \, \left( {\tilde M}_{i,+}^{({\tau})}{\Sigma}_{i}^{(-{\tau})} +
{\tilde M}_{i,-}^{(-{\tau})}{\Sigma}_{i}^{({\tau})} \right) + \sum_{{\tilde e}\in {\tilde{\Eg}}} \, e^{{\tau}( {\tilde m}_{\tilde e} + {\ve}) } \, {\Sigma}_{s({\tilde e})}^{({\tau})}{\Sigma}_{t({\tilde e})}^{(-{\tau})} \right\} \right] 
\label{eq:instpf2}
\end{multline}
Here we separated the contributions of the chiral multiplets whose bosonic components remain scalars on $C$ and the chiral multiplets (whose multiplicity spaces and masses have  tildes) which become $1$-forms. In order to preserve the topological supersymmetry 
 the superpotential $W$ (which comes from the cycles in the quiver) is a $(1,0)$-form on the worldsheet (not to be confused with the twisted superpotential $\tilde{W}$) . For example, in the ${\CalN}=(4,4)$ case it has the form:
 \beq
 W =  {\mu}_{\BC} ({\Phi}_{z}) dz
 \eeq 
The possible twists (such as the one making the adjoint chiral $\Phi$ a one-form $\Phi_z$) are encoded in the $\ve$-dependence of the (anti-)fundamental masses in $M_{i,\pm}$ and the bi-fundamental masses $m_e$. 

$\bullet$ An important remark is in order. It would appear that
the partition functions \eqref{eq:instpf}, \eqref{eq:instpf2}
depend on both the Coulomb parameters $a_{i, \alpha}$ and the masses $m_e$, $m_{i, f}$. However, in two dimensions one does not expect to be able to freely fix the scalars in the 
vector multiplet. Instead, they are dynamically fixed by the choice of the (generically massive) vacuum the theory falls in the infrared. In practice this means that one actually should set
$a_{i, \alpha} = a_{i, \alpha}(m)$ to be equal to  the classical values determined by the twisted masses. These are discrete choices, which correspond to the choice of the equivariant cohomology class of the effective target space.

\subsection{Twisted superpotential and supersymmetric ground states}

The vacua of the ${\CalN}=(2,2)$ theory can be then determined by finding a saddle point in the sum over the fluxes
in the limit ${\ve} \to 0$, as is done in four dimensions 
\cite{NS3, NO}:
\beq
{\CalZ}_{2d}^{{\CalN}=2} \sim {\exp}\, \frac{1}{\ve} {\tilde\CalW} ({\sigma}) + \ldots 
\label{eq:czasy}
\eeq
where ${\sigma}_{i,\alpha}$ are 
found from the Bethe-like equations 
\beq
\frac{P_{i,+}({\sigma}_{i, \alpha})}{P_{i,-}({\sigma}_{i,\alpha})}
\frac{{\tilde P}_{i,+}({\sigma}_{i, \alpha})}{{\tilde P}_{i,-}({\sigma}_{i,\alpha})}  \frac{\prod\limits_{e \in t^{-1}(i)} Q_{s(e)}({\sigma}_{i, \alpha} - m_{e})}{\prod\limits_{e \in s^{-1}(i)} Q_{t(e)}({\sigma}_{i, \alpha} + m_{e})} \frac{\prod\limits_{{\tilde e} \in t^{-1}(i)} Q_{s({\tilde e})}({\sigma}_{i, \alpha} - {\tilde m}_{\tilde e})}{\prod\limits_{{\tilde e} \in s^{-1}(i)} Q_{t({\tilde e})}({\sigma}_{i, \alpha} + {\tilde m}_{\tilde e})} = {\qe}_{i}
\label{eq:bae}
\eeq
where $e \in \Eg$, ${\tilde e} \in \tilde{\Eg}$, 
for all ${\alpha} = 1, \ldots , n_{i}$, and
\beq
 P_{i, \pm}(x) = \prod_{f=1}^{m_{i, \pm}} (x - m_{\pm, f}) \, , \ Q_{i}(x) = \prod_{{\alpha}=1}^{n_{i}} ( x- {\sigma}_{i,\alpha} )\, ,  \
 {\tilde P}_{i, \pm}(x) = \prod_{f=1}^{{\tilde m}_{i, \pm}} (x - {\tilde m}_{\pm, f}) \, .
\label{eq:pqpol}
\eeq
The classical values $\left( a_{i,\alpha}(m) \right)$ are the poles of the meromorphic form:
\beq
\bigwedge_{i, \alpha} d x_{i,\alpha} \ \prod_{i \in \Vg} \prod\limits_{{\alpha} \neq {\beta}} (x_{i,\alpha} - x_{i, \beta})\prod\limits_{\alpha = 1}^{n_{i}} \frac{{\tilde P}_{i,+}(x_{i,\alpha}) {\tilde P}_{i, -} (x_{i,\alpha})}{P_{i,+}(x_{i,\alpha})P_{i,-}(x_{i, \alpha})} \  \frac{\prod\limits_{{\tilde e} \in \tilde\Eg}  \prod\limits_{{\alpha} \in [n_{s({\tilde e})}]} \prod\limits_{\beta \in [n_{t({\tilde e})}]} \left( x_{s({\tilde e}), \alpha} - x_{t({\tilde e}), \beta} + {\tilde m}_{{\tilde e}} \right)}{\prod\limits_{e \in \Eg}  \prod\limits_{{\alpha} \in [n_{s(e)}]} \prod\limits_{\beta \in [n_{t(e)}]} \left( x_{s(e), \alpha} - x_{t(e), \beta} + m_{e} \right)} 
\eeq

Let us recall the effective twisted superpotential for the softly broken ${\CalN}=(4,4)$ theories
\begin{multline}
{\tilde{\CalW}} \left( {\sigma} ; {\mu} ; u \right) =  
\sum\limits_{i \in \Vg}\, \sum\limits_{{\alpha} \in [n_{i}]} \ \left(\, {\log}({\qe}_{i})\, {\sigma}_{i,\alpha}+ 
\sum\limits_{{\beta} \in [n_{i}]}
{\varpi} 
\left( -2u + {\sigma}_{i,\alpha} - {\sigma}_{i,\beta} \right)
+ \right. \\
\left. \qquad\qquad\qquad\qquad\qquad\qquad + \quad \sum\limits_{{\sf f} \in [m_{i}]} \,
\left( {\varpi} \left(  u + {\sigma}_{i,\alpha}- m_{i, \sf f}\right) +
{\varpi} \left( u - {\sigma}_{i,\alpha} + m_{i, \sf f} \right)   \right) \right)
\ \\
+ \sum\limits_{e \in \Eg} \, \sum\limits_{{\alpha} \in [n_{t(e)}]} \, \sum\limits_{{\beta} \in [n_{s(e)}] } \ 
\left( {\varpi} \left( u + m_{e}+ {\sigma}_{t(e), \alpha} - {\sigma}_{s(e), \beta}  \right)
+ {\varpi} \left( u - m_{e} + {\sigma}_{s(e), \beta}  - {\sigma}_{t(e), \alpha} \right) \right)\
\end{multline}
where 
\beq
{\varpi}(z) = z\, \left( \, {\log}(z) - 1 \, \right) \, , \qquad {\exp} \, {\varpi}^{\prime}(z) = z
\label{eq:1loop2d}
\eeq
The equations \eqref{eq:bae} describe the generalized critical points of $\tilde{\CalW}$:
\beq
{\exp}\, \frac{{\partial}{\tilde{\CalW}}}{{\partial}{\sigma}_{i,\alpha}} = 1
\eeq
$\bullet$ Note that our partition functions are the sums over discrete sets. Since these partition functions are obtained by the equivariant integration over the moduli spaces which can be given a finite-dimensional K\"ahler quotients, they can also be given a contour integral expression a la \cite{MNS}. The asymptotics \eqref{eq:czasy} then becomes the analogue of the critical level limit of the integral representation (free field) of the solutions 
to KZ equations \cite{Dotsenko:1984nm, Gerasimov:1990fi, Awata:1991, Babujian:1993tm, Babujian:1993ts, Tarasov:1993vs, Frenkel:1994be}.  The connection to Bethe ansatz equations is at first surprising and forms the basis of the Bethe/gauge-correspondence \cite{NS1, NS2, NS3}. Its mathematical
foundations are laid out in \cite{Maulik:2012wi}.

\subsection{Sigma models on grassmanians and vector bundles over them}

The following examples are important:

$\bullet$ The Grassmanian $Gr(n,m)$ sigma model: it is the $A_1$ type theory, with the quiver consisting of one node, with ${\bM}_{+} = {\BC}^{m}$, ${\bN} = {\BC}^{n}$, ${\bM}_{-} = 0$, 
\beq
M_{+}^{({\tau})} = \sum_{f=1}^{m} e^{{\tau}m_{f}} .
\eeq 
A simple computation shows
\beq
\frac{{\CalZ}_{2d}}{{\CalZ}_{2d}^{\rm pert}} = \sum\limits_{d_{1}, \ldots , d_{n} \geq 0}\ {\qe}^{d_1 + d_2 + \ldots + d_{n}} \ J_{\vec d}^{Gr}
\label{eq:grpf}
\eeq
where 
\begin{multline}
J^{Gr}_{\vec d} = {\epsilon} \left[  \frac{( M_{+}^{({\tau})} - {\Sigma}^{({\tau})} ){\Sigma}^{(-{\tau})} - ( M_{+}^{({\tau})} - N^{({\tau})} )N^{(-{\tau})}}{1-e^{{\tau}{\ve}}}
\right] = \\
\prod_{1\leq i < j \leq n} \frac{a_{i} - a_{j} + {\ve} (d_{i}  - d_{j})}{a_{i}- a_{j}} \, \prod_{f=1}^{m} \prod_{i=1}^{n} \prod_{l=1}^{d_{i}} \frac{1}{a_{i} - m_{f} + {\ve} l} 
\label{eq:grpf2}
\end{multline}
which agrees with \cite{Bertram:2003qd}  (with $\hbar = - \ve$) provided we view the partition function \eqref{eq:grpf} as valued in the $U(m)$-equivariant cohomology of $Gr(n,m)$, which is the ring of
$S(n)$-invariant polynomials in $a_{1}, \ldots , a_{n}$ subject to the relation 
\beq
A(x) B(x) = P(x) = \prod_{f=1}^{m} (x - m_{f}) 
\label{eq:abp}
\eeq
where $A(x) = \prod_{i=1}^{n}(x - a_{i})$ and the degree $m-n$ polynomial $B(x)$ is determined from \eqref{eq:abp} (see also
\cite{Hori:2000kt} for the period interpretation of \eqref{eq:grpf2} in the
context of mirror symmetry). As the ${\BC}[m_{f}]^{S(m)}$-module the ring has the rank 
$\left( \begin{matrix} m \\  n \\ \end{matrix} \right)$. 

$\bullet$ The next example is the gauged linear sigma model
corresponding to the vector bundle $F$ over $Gr(n,m)$, with the
fiber ${\rm Hom}( {\CalE}, {\bM}_{-}) $ being $k$ copies of the  dual to the tautological rank $n$ bundle ${\CalE}^{\vee}$ over
$Gr(n,m)$. In our conventions this model corresponds to 
${\bM}_{+} = {\BC}^{m}$, ${\bM}_{-} = {\BC}^{k}$. 
Let
\beq
\begin{aligned}
& M_{+}^{({\tau})} = \sum_{f=1}^{m} e^{{\tau}m_{+,f}}\, , \qquad
M_{-}^{({\tau})} = \sum_{f=1}^{k} e^{{\tau}m_{-,f}} \, , \\
& P_{+}(x) = \prod_{f=1}^{m} ( x- m_{+,f}) \, , \qquad P_{-}(x) = \prod_{f=1}^{k} ( x - m_{-, f} ) \\
\end{aligned}
\label{eq:mpmm}
\eeq
The instanton partition function evaluates to (up to the overall
sign and a possible redefinition $\qe \to - \qe$):
\beq
\frac{{\CalZ}_{2d}}{{\CalZ}_{2d}^{\rm pert}} = \sum\limits_{d_{1}, \ldots , d_{n} \geq 0}\ {\qe}^{d_1 + d_2 + \ldots + d_{n}} \ J_{\vec d}^{F}
\label{eq:grpf3}
\eeq
with 
\beq
J^{F}_{\vec d} = 
\prod_{1\leq i < j \leq n} \frac{a_{i} - a_{j} + {\ve} (d_{i}  - d_{j})}{a_{i}- a_{j}} \, \prod_{i=1}^{n} \prod_{l=1}^{d_{i}} \frac{P_{-}\left( a_{i} + {\ve}(l-1) \right)}{P_{+} (a_{i}  + {\ve} l)} 
\label{eq:grpf4}
\eeq
where $a_{1} \neq a_{2} \neq \ldots \neq a_{n}$ are to be chosen from the roots of $P_{+}$. 

$\bullet$ The general Nakajima varieties, which are the hyperk\"ahler quotients of the space of quiver representations
will be discussed in this framework in our future work. We plan to connect our approach to the beautiful results of \cite{Braverman:2010ei}, \cite{Aganagic:2017gsx} and especially \cite{Maulik:2012wi}. 

\section{Noncommutative description}

In this section we explain how one can derive our formula \eqref{eq:instpf} (as well as the formulae for the four dimensional instanton partition functions \cite{N2,N3,N4})  using noncommutative gauge theory, which is, in some sense, a minimal model
of the open string field theory \cite{SW3}. The gauged linear sigma model
for an affine ADE type quiver
can be realized as the low energy limit of the theory of open strings 
ending on a collection of $D$-branes wrapping various cycles on an ADE
singularity, and two noncompact directions transverse to it. The equivariant parameters
$u$ and $\ve$ are realized geometrically. The full ten dimensional geometry is 
$S_{ADE} \times B_{u} \times D_{\ve} \times T^{2}$
where
$B_{u} \approx {\BC}$, $D_{\ve} \approx {\BC}$ are fibered holomorphically over $T^2 = E$, with complex
parameters $u$ and $\ve$ being simply the points on the Jacobian $Jac(E)$. The factor
$S_{ADE}$ is the complex surface which is a resolution of singularities of 
$L \otimes {\BC}^{2}/{\Gamma}_{\gamma}$,  with $L^{\otimes -2} \approx B_{u} \otimes D_{\ve}$ 
as line bundles over $E$, where $\Gamma_{\gamma} \subset SU(2)$ is a finite subgroup, whose
representations are encoded by the quiver. 

By turning an appropriate $B$-field on $S_{ADE} \times B_{u} \times D_{\ve}$ and taking the Seiberg-Witten limit \cite{SW3} one arrives at the theory which can be roughly described as
the dimensional reduction of the ${\CalN}=1$,$d=10$ super-Yang-Mills theory down to two dimensions (with the worldsheet being $T^2$), with a peculiar gauge group: a group of unitary transformations of a Hilbert space $\CalH$. This space is described as follows. 

Let ${\CalH}_{2}$ be the Fock space of states of 
two harmonic oscillators:
\beq
{\CalH}_{2} = {\BC}[{\ba}_{1}^{\dagger}, {\ba}_{2}^{\dagger}]\, \vert \, {\rm vac} \rangle
\label{eq:fock2}
\eeq
with the operators ${\ba}_{1}, {\ba}_{2}, {\ba}_{1}^{\dagger}, {\ba}_{2}^{\dagger}$
obeying the Heisenberg algebra
$[{\ba}_{i}, {\ba}_{j}^{\dagger}] = {\delta}_{ij}$, $i,j = 1, 2$, $[{\ba}_{1}, {\ba}_{2}] = 0$. 

Let $\4$, $\6$ be the sets defined in \cite{N5}. For $a \in \4$, $A \in \6$, such that $a \in A$  define $i(a, A)$ to be $1$ if $A = ab$ with $a<b$ and $2$ if $a>b$. Fix six Hermitian vector spaces $N_A$, $A \in \6$. 

Let ${\CalH} \approx {\bN} \otimes {\CalH}_{2}$, with 
${\bN} = \bigoplus_{A} \, N_{A}$, $A \in \6$. Define
\beq
A_{a} = \frac{1}{\sqrt{2}} \, \bigoplus_{A \in \6\, , \ a \in A} 1_{N_{A}} \otimes 
{\ba}_{i(a,A)} 
\eeq
In other words, we have distributed two creation operators between six copies of the Fock space tensored with some finite-dimensional vector space. 
Then 
\beq
[A_{a}, A_{b} ] = 0, \qquad [ A_{a}, A_{b}^{\dagger} ] = \frac 12 {\delta}_{ab} \bigoplus\limits_{A \in \6, A \ni a} 1_{N_{A}}
\label{eq:aaab}
\eeq 
Now let us look for the quadruples of operators $Z_{a}$, $a \in \4$, obeying a weaker version of \eqref{eq:aaab}:
\beq
\begin{aligned}
&  [ Z_{a}, Z_{b} ] + {\ve}_{abcd} [Z_{c} , Z_{d}]^{\dagger} = 0 \, , \\
& \qquad \sum_{a\in \4} [Z_{a}, Z_{a}^{\dagger}] \, = \, 1_{\CalH} \\
\end{aligned}
\label{eq:ncinst}
\eeq 
which are close, in the approprate sense, to the basic solution $Z_a = A_a$. We identify solutions which differ by the gauge transformation $Z_a \mapsto g^{-1} Z_a g$, $g \in U({\CalH})$. 

Now let us count the solutions \eqref{eq:ncinst} equivariantly with respect to the maximal torus ${\BT}$ of $SU(4)$ which acts linearly on the $Z_a$'s:
\beq
Z_{a} \mapsto \sum_{b \in \4} u_{a}^{b} Z_{b} \, \qquad u u^{\dagger} = 1, \qquad det(u) = 1
\label{eq:rotsym}
\eeq
The basic solution is invariant under \eqref{eq:rotsym} in the sense that such a transformation can be undone by the unitary similarity transformation. Infinitesimally:
\beq
{\ve}_{a} A_{a} = [ {\Phi}_{0} , A_{a} ]
\eeq
where
\beq
{\Phi}_{0}  = \bigoplus\limits_{a<b \in \4} {\rm diag} ({\ac}_{ab, \alpha})_{{\alpha} \in [n_{ab}]} \otimes 1_{{\CalH}_2} + 
1_{N_{ab}} \otimes \left( {\ve}_{a} {\bf a}^{\dagger}_{1}{\bf a}_{1} + {\ve}_{b} {\bf a}^{\dagger}_{2} {\bf a}_{2} \right) 
\eeq 
The infinite-dimensional version of the fixed point localization technique would reduce the counting to the enumeration of the solutions which are invariant, modulo gauge transformations:
\beq
{\ve}_{a} Z_{a} = [ {\Phi}, Z_{a} ]
\eeq
where  $\Phi - \Phi_{0}$ is a compact self-adjoint operator in $\CalH$, with the 
eigenvalues of the form ${\ac}_{ab, {\alpha}} + {\ve}_{a} i_{a} + {\ve}_{b}i_{b}$, with
$i_{1}, i_{2}, i_{3}, i_{4} \in {\BZ}$, ${\alpha}  = 1, \ldots , n_{ab} = {\rm dim}N_{ab}$. 

The fixed point contribution, including the perturbative prefactor is given by the ratio of regularized determinants:
\beq
\frac{{\rm Det}' \left( {\rm ad}({\Phi}) \right) \, {\rm Pf} \left( \prod\limits_{a< b} \,  \left( {\rm ad}({\Phi}) + {\ve}_{a} + {\ve}_{b} \right)\right)}{\prod\limits_{a} \, {\rm Det} \left( {\rm ad}({\Phi}) + {\ve}_{a}  \right)} \sim {\epsilon}\left[ - P_{123} H H^{*} \right] \ , \eeq
where $P_{123} = (1-q_{1})(1-q_{2})(1-q_{3})$, $q_{a} = e^{{\tau}{\ve}_{a}}$, $H = {\Tr}_{\CalH} e^{\tau\Phi}$, $H^{*} = {\Tr}_{\CalH} e^{-\tau\Phi}$, etc. 
To arrive at \eqref{eq:instpf} one imposes additional ${\Gamma}_{\gamma}$-orbifold projections, as in \cite{N4, N5}.

\section{Three types of surface defects}

In this section we discuss three types of surface defects: the ones defined using the orbifold
of the spacetime by a cyclic group; the ones obtained by coupling the theory to the
nearly Higgsed gauge theory, we call it a quiver construction for the reasons which will become clear below; and the folded defects, which correspond to the gauge theory on singular spacetime, a product of a smooth Riemann surface and a nodal curve. The connection between the first two types of defects has been conjectured long time ago, and recently has been discussed in the context of the renormalization theory in \cite{Frenkel:2015rda}. The rigorous results were obtained in \cite{Jeong:2018qpc} leading to the proof of the conjectures of \cite{Nekrasov:2011bc}. 

\subsection{The orbifold construction}

The operators defined by the orbifold projection were studied earlier
in \cite{AT, Ashok:2017lko, Awata:2010bz, Bullimore:2014awa, Nawata:2014nca}. We give the geometric realization of these operators below.  

We start by recalling the
ADHM construction: consider the moduli space ${\mM}_{k,n}$ of matrices $(B_{1}, B_{2}, I, J)$, $B_{1}, B_{2} \in {\rm End}(K), I \in {\rm Hom}(N, K), J \in {\rm Hom}(K, N)$, obeying the following equations and stability conditions: 
\beq
\begin{aligned}
& [ B_{1}, B_{2} ] + IJ = 0  \\
& {\rm Stability}: \ {\BC} [B_{1}, B_{2}] \, I(N) \ = \ K \\
\end{aligned}
\label{eq:adhm}
\eeq
modulo the symmetry  \beq (B_{1}, B_{2}, I, J) \longrightarrow (g^{-1}B_{1}g, g^{-1} B_{2}g , g^{-1}I, J g)\, , \qquad g \in GL(K) \eeq

The supersymmetric partition function of ${\CalN}=2$ supersymmetric gauge theory on ${\BR}^{4} \approx {\BC}^{2}$ with appropriate supersymmetric boundary conditions at infinity, with the gauge group $U(n)$, in the instanton sector $k$,  localizes onto the integral of an equivariant differential form on ${\mM}_{k,n}$.  More precisely, in defining ${\mM}_{k,n}$
we used a particular stability condition in \eqref{eq:adhm}. It actually corresponds to deforming ${\BR}^{4}$ into a noncommutative space ${\BR}^{4}_{\theta}$, such that the holomorphic coordinates $z_{1}, z_{2}$ remain commuting, while their commutators with the conjugates obey: $[z_{1}, {\zb}_{1}] + [z_{2}, {\zb}_{2}] = - {\zeta}$, ${\zeta} > 0$. 

Now let us insert a surface operator along the ${\BR}^{2} \approx {\BC}^{1}$ plane, with the coordinate $z_{1}$, i.e. the surface is defined by the equation $z_{2} = 0$. 
  
Motivated by \cite{B, Biswas, FK}  
we perform the orbifold, following the \cite{DM} construction: make $N$ and $K$ the ${\BZ}_{p}$-modules:
\beq
N = \bigoplus_{\bnu} N_{\bnu} \otimes {\CalR}_{\bnu} \, , \quad K = \bigoplus_{\bnu} K_{\bnu} \otimes {\CalR}_{\bnu} 
\label{eq:orbkn}
\eeq
with ${\CalR}_{\bnu}$, $\bnu = 0, 1, \ldots, p-1$, ${\bnu} + p \equiv \bnu$, being the one-dimensional irreducible representation of ${\BZ}_{p}$, where the generator ${\omega}$ acts via:
\beq
T_{{\CalR}_{\bnu}} ({\omega} ) = {\varpi}^{\bnu} \equiv e^{\frac{2\pi\ii \bnu}{p}}
\eeq
and impose the equivariance condition, with ${\Omega}_{K} = T_{K}({\omega}), \, {\Omega}_{N} = T_{N}({\omega})$:
\beq
B_{1} = {\Omega}^{-1}_{K} B_{1} {\Omega}_{K}\, , \qquad {\varpi} B_{2} =  {\Omega}_{K}^{-1} B_{2} {\Omega}_{K} \, , \qquad 
I = {\Omega}^{-1}_{K} I {\Omega}_{N} \, \qquad {\varpi} J = {\Omega}_{N}^{-1} J {\Omega}_{K} \eeq
which imply that the operators $B_{1}, B_{2}, I, J$ decompose as $B_{1} = (B_{1,{\bnu}})_{\bnu}$, $B_{2} = (B_{2, {\bnu}})_{\bnu}$, $I = (I_{\bnu})_{\bnu}$, $J = (J_{\bnu})_{\bnu}$, with 
\beq
B_{1, \bnu} : K_{\bnu} \to K_{\bnu}\, , \ B_{2, \bnu}: K_{\bnu} \to K_{{\bnu}-1}
\eeq 
\beq
I_{\bnu}: N_{\bnu} \to K_{\bnu}\, , \ J_{\bnu}: K_{\bnu} \to N_{{\bnu}-1}
\eeq
The projected equations
\beq
B_{1, \bnu - 1} B_{2, \bnu} - B_{2, \bnu} B_{1, \bnu} + I_{\bnu - 1} J_{\bnu} = 0
\eeq
Now use the identity:
\beq
[ B_{1}, B_{2}^{p} ] = \sum_{m=0}^{p-1} B_{2}^{m} [ B_{1}, B_{2} ]  B_{2}^{p-1-m} = - 
\sum_{m=0}^{p-1} B_{2}^{m} IJ  B_{2}^{p-1-m}
\eeq
to arrive at:
\beq
[ {\tilde B}_{1}, {\tilde B}_{2} ] + {\tilde I} {\tilde J} = 0
\eeq
where ${\tilde B}_{1}, {\tilde B}_{2} \in {\rm End}({\tilde K}), {\tilde I} \in {\rm Hom}({\tilde N}, {\tilde K}), {\tilde J} \in {\rm Hom}({\tilde K}, {\tilde N})$ are the new ADHM matrices, with
\beq
\begin{aligned}
& {\tilde K} \ = \ K_{p-1}\, , \qquad {\tilde N} \ = \ \bigoplus_{\bnu}\ N_{\bnu} \, ,  \\
& {\tilde B}_{1} = B_{1, p-1} \, , \qquad {\tilde B}_{2} = B_{2,0} B_{2,1} \ldots B_{2,p-1}\, ,  \\
& {\tilde I} = \sum_{m=0}^{p-1} B_{2,0} B_{2,1} \ldots B_{2,m-1} I_{m-1} \, , \qquad
{\tilde J} =   \sum_{m=0}^{p-1} J_{m} B_{2,m+1} \ldots B_{2,p-1} \, , \\
\end{aligned}
\eeq
We thus get a map: ${\mM}_{k,n}^{\rm orb} \to {\mM}_{{\tilde k}, n}$. Integrating along the fibers of this map we produce a cohomology class of $\mM_{{\tilde k}, n}$ which one can express in terms of the equivariant Donaldson classes. 

Let us now give the explicit expression of this class, the representative of the corresponding surface operator in the $\CalQ$-cohomology of the gauge theory in the $\Omega$-background. We do it in two examples, the $A_1$ theory, i.e. the $U(N)$ theory with $2N$ fundamental hypermultiplets, and
the ${\CalN}=2^{*}$ theory. 

In the both cases the first step is to perform the ${\BZ}_{p}$-projection of the vector multiplet contribution to the instanton measure.  To save the space we omit the explicit mention of the $\tau$-dependence, and denote by $*$ the $\tau \to -\tau$ conjugation, as well as ${\CalR}_{\bnu}^{*} = {\CalR}_{p-\bnu}$:
\beq
{\epsilon} \left[ - \sum_{{\bnu}, {\bnu}''=0}^{p-1} \frac{S_{\bnu''} S_{\bnu}^{*}}{P_{1} (1- {\tilde q}_{2})} \sum_{{\bnu}'=0}^{p-1} q_{2}^{{\bnu}'} {\delta}_{\bnu''+\bnu' - \bnu}^{{\BZ}_{p}} \right]
\label{eq:projme}
\eeq
where $P_{1} = 1- q_{1}$, ${\tilde q}_{2} = q_{2}^{p}$ and we used the identity:
\beq
\frac{1}{1-q_{2} {\CalR}_{1}} = \frac{1}{1- {\tilde q}_{2}}\sum_{{\bnu}=0}^{p-1} q_{2}^{\bnu} {\CalR}_{\bnu}
\eeq
in $K[{\BZ}_{p}]\otimes {\BC}$. 
Define 
\beq
{\tilde S}_{\bnu} = S_{\bnu} q_{2}^{-\bnu} = {\tilde N}_{\bnu} - 
P_{1} {\tilde K}_{\bnu} + P_{1}  \begin{cases} {\tilde K}_{{\bnu}-1} \, , \ {\bnu} > 0 \\
{\tilde K}_{p-1} {\tilde q}_{2} \, , \ {\bnu} = 0 \end{cases}\ ,
\label{eq:snk}
\eeq 
\beq
{\tilde S} = \sum_{{\bnu}=0}^{p-1} {\tilde S}_{\bnu} = 
{\tilde N} - P_{1}{\tilde P}_{2} {\tilde K}
\label{eq:effs}
\eeq
where ${\tilde P}_{2} = 1 - {\tilde q}_{2}$. 
Now we can rewrite \eqref{eq:projme} as
\beq
{\epsilon}\left[ - \frac{{\tilde S}{\tilde S}^{*}}{P_{1}{\tilde P}_{2}} \right]
\cdot {\epsilon} \left[  \frac{1}{P_{1}}\ \sum\limits_{0 \leq {\bnu}' < {\bnu}'' < p} {\tilde S}_{\bnu''} {\tilde S}_{\bnu'}^{*} \right]
\eeq
The first factor is the usual four dimensional vector multiplet instanton measure, while the second is the two dimensional gauged linear sigma model contribution, corresponding to the sigma model on the (partial) flag variety, cf.
\eqref{eq:instpf2}. 

Let us clarify the latter statement. At the level of fixed points, the $N$-tuples of Young diagrams ${\lambda}^{({\alpha})}$, ${\alpha}  = 1, \ldots , N$ enumerate the fixed points of the $U(1)^{N} \times U(1) \times U(1)$-action both on ${\mM}_{k,n}^{\rm orb}$ and on ${\mM}_{{\tilde k}, n}$. However, the map is many-to-one. 
Let $c: [N] \to \{ 0, 1, \ldots , p-1 \}$ be the coloring function, which describes the
decomposition \eqref{eq:orbkn}: 
\beq
N_{\bnu} = \sum\limits_{\alpha \in  c^{-1}({\bnu})} \, e^{{\tau}a_{\alpha}} 
\eeq
Then
\beq
K_{\bnu} \ = \ \sum_{{\alpha}=1}^{N} e^{{\tau}a_{\alpha}} \qquad
\sum_{i =1}^{\left( {\lambda}^{({\alpha})} \right)^{t}} q_{1}^{i-1} \ \sum\limits_{1 \leq j \leq {\lambda}^{({\alpha})}_{i}\, , \ j-1+c({\alpha}) \equiv \bnu (p)} q_{2}^{j-1}
\eeq
Let
\beq
{\tilde a}_{\alpha} = a_{\alpha} - {\ve}_{2} c({\alpha})\, , 
\eeq 
so that
\beq
{\tilde N}_{\bnu} = \sum_{{\alpha} \in c^{-1}({\bnu})} e^{{\tau}{\tilde a}_{\alpha}} \, , 
\eeq
\begin{multline}
{\tilde K}_{\bnu} = K_{\bnu} q_{2}^{-\bnu} = 
\sum_{{\alpha}=1}^{N} e^{{\tau}{\tilde a}_{\alpha}}
\sum_{i =1}^{\left( {\lambda}^{({\alpha})}\right)^{t}} q_{1}^{i-1} \ \sum\limits_{1 \leq j \leq {\lambda}^{({\alpha})}_{i}\, , \ j-1+c({\alpha}) \equiv \bnu (p)} {\tilde q}_{2}^{\frac{j-1+c({\alpha})-\bnu}{p}} = \\
\sum_{{\alpha}=1}^{N} e^{{\tau}{\tilde a}_{\alpha}}
\sum_{i =1}^{\left( {\lambda}^{({\alpha})}\right)^{t}} q_{1}^{i-1} \ \sum\limits_{j=1\, {\tiny or} \, 2}^{l_{{\alpha},i,{\bnu}}} {\tilde q}_{2}^{j-1} 
\end{multline}
where $l_{{\alpha},i,{\bnu}} = \left[ \frac{{\lambda}^{({\alpha})}_{i}+ c({\alpha}) - {\bnu} + p-1}{p} \right]$, and the lower limit of the sum over $j$ is equal to $1$
for $c({\alpha}) \leq \bnu$ and $2$ otherwise. 
We can also write, in terms of the dual partitions ${\lambda}^{({\alpha})t}$:
\beq
P_{1}{\tilde K}_{\bnu} = \sum_{{\alpha}=1}^{N} \, e^{{\tilde a}_{\alpha}} \, \sum_{J = 1\,  {\tiny or} \, 2} {\tilde q}_{2}^{J-1} \left( 1 - q_{1}^{{\lambda}_{1 - c({\alpha})+ p(J-1)+ \bnu}^{({\alpha})t}} \right) 
\eeq 
where, again, $J \geq 1$ when ${\bnu} \geq c({\alpha})$, and
$J \geq 2$ when ${\bnu} < c({\alpha})$.

Now, for ${\bnu} = p-1$:
\beq
{\tilde K} = \sum_{{\alpha}=1}^{N} e^{{\tau}{\tilde a}_{\alpha}}
\sum_{i =1}^{\left( {\Lambda}^{({\alpha})}\right)^{t}} q_{1}^{i-1} \ \sum\limits_{j=1}^{{\Lambda}_{i}^{({\alpha})}} {\tilde q}_{2}^{j-1}
\eeq
where
\beq
{\Lambda}^{({\alpha})}_{i} = \left[ \frac{{\lambda}_{i}^{({\alpha})} + c({\alpha})}{p} \right]
\label{eq:rhom}
\eeq
We denote by $\rho_{c}$ the map of the set of $N$-tuples
of partitions to itself, given by \eqref{eq:rhom}:
\beq
\left( {\Lambda}^{({\alpha})} \right) = {\rho}_{c} \left( 
{\lambda}^{({\alpha})} \right) 
\label{eq:rhom2}
\eeq
The map $\rho_{c}$ is, in general, $\infty : 1$. Indeed, 
any ${\lambda}_{i}^{({\alpha})} < p - c({\alpha})$ for sufficiently large $i$ will map to ${\Lambda}_{i}^{({\alpha})} = 0$. There is no room for such ambiguity for ${\alpha} \in c^{-1}(p-1)$. 
  
Let
\beq
k_{\bnu} = | K_{\bnu} | \ = \ \sum_{{\alpha}=1}^{N} \sum_{i=1}^{({\lambda}^{({\alpha})})^{t}} \left( \ \left[ \frac{{\lambda}^{({\alpha})}_{i}+ c({\alpha}) - {\bnu} + p-1}{p} \right]   - \left[ \frac{c({\alpha}) - {\bnu} + p-1}{p} \right] \, \right) 
\eeq
In particular
\beq
{\tilde k} = k_{p-1} = \sum_{{\alpha}=1}^{N} | {\Lambda}^{({\alpha})} |
\label{eq:instch4d}
\eeq
Define the auxiliary variables $(z_{\bnu})$ and $\qe$, via
\beq
{\qe}_{0} = z_{1}/z_{0}\, , \ {\qe}_{1} = z_{2}/z_{1} \, , \ \ldots \, , \ {\qe}_{p-1} = {\qe}\, z_{0}/z_{p-1}
\label{eq:zfrq}
\eeq 

Define ${\Sigma}_{0} = 0$, ${\Sigma}_{p} = {\tilde S}$
and ${\Sigma}_{\bnu}$, ${\bnu} = 1, \ldots , p-1$, via:
\beq
{\tilde S}_{\bnu} = {\Sigma}_{{\bnu}+1} - {\Sigma}_{\bnu}
\label{eq:sfrsig}
\eeq
so that
\beq
{\Sigma}_{\bnu} = {\tilde N}_{0} + {\tilde N}_{1} + \ldots + {\tilde N}_{\bnu -1} - P_{1}{\tilde K}_{\bnu -1} + {\tilde q}_{2} P_{1}{\tilde K} \ .
\eeq
Thus, the surface defect, as an operator of the four dimensional chiral ring, is given by:
\beq
{\CalI}_{{\Lambda}^{({\alpha})}} = 
\sum_{\left( {\lambda}^{({\alpha})} \right) \in {\rho}_{c}^{-1}({\Lambda}^{({\alpha})})} \ \ \prod_{\bnu = 0}^{p-1}
\ z_{\bnu}^{k_{\bnu -1} - k_{\bnu}} \ \cdot \ 
{\epsilon} \left[ \frac{1}{P_{1}}\ \sum\limits_{{\bnu}=1}^{p-1} \left( {\Sigma}_{\bnu +1} - {\Sigma}_{\bnu} \right) {\Sigma}_{\bnu}^{*} \right]
\eeq
This is the partition function of the two dimensional gauged linear sigma model on the partial flag variety $Flags(d_{1}, d_{2}, \ldots, N)$, with
\beq
d_{\bnu} = \# \, \{ \ {\alpha}\ | \ c({\alpha}) < {\bnu} \ \} \, , 
\label{eq:dbnu}
\eeq 
coupled to the four dimensional gauge theory, via $\Sigma_{p} = {\tilde S}$. 

This would be the end of the story for the pure super-Yang-Mills theory. In particular, for regular defect, $p = N$, with $d_{\bnu} = {\bnu}$, we get the setup of \cite{Braverman:2004vv}. However, the power of our approach is the ability to include the matter fields. In the $A_1$ case the orbifold projection must also include the decomposition of the matter multiplet:
\beq
M \ = \ \bigoplus_{{\bnu}=0}^{p-1} \, M_{\bnu} \otimes {\CalR}_{\bnu}
\label{eq:mbnu}
\eeq
The additional factor in the instanton measure is
\beq
{\epsilon}\left[ - \sum_{\bnu} {\tilde M}_{\bnu}^{*} {\tilde K}_{\bnu} \right] = {\epsilon} \left[ - {\tilde M}^{*}{\tilde K} \right] \cdot  {\epsilon}
\left[ - \frac{1}{P_{1}} \sum_{{\bnu}=0}^{p-1} {\tilde M}_{\bnu}^{*} \left( {\tilde N}_{0} + {\tilde N}_{1} + \ldots + {\tilde N}_{\bnu }  - {\Sigma}_{\bnu+1} \right) \right]
\eeq
where 
\beq
{\tilde M} = {\tilde q}_{2}^{-1} \sum_{\bnu} {\tilde M}_{\bnu}
\eeq
which means that the surface operator is now the partition function of the two dimensional sigma model with the target space the total space of the vector bundle over
$Flags(d_{1}, d_{2}, \ldots , d_{p-1}, N)$ with the fiber
\beq
\bigoplus\limits_{i=1}^{p-1}\ {\rm Hom}({\CalE}_{i}, {\tilde M}_{i-1})
\eeq
where ${\CalE}_{i}$, $rk{\CalE}_{i} = d_{i}$ is the $i$'th tautological bundle. 

In the ${\CalN}=2^{*}$ case the instanton measure
is 
\beq
{\epsilon} \left[ - P_{3} \sum_{{\bnu}, {\bnu}''=0}^{p-1} \frac{S_{\bnu''} S_{\bnu}^{*}}{P_{1} (1- {\tilde q}_{2})} \sum_{{\bnu}'=0}^{p-1} q_{2}^{{\bnu}'} {\delta}_{\bnu''+\bnu' - \bnu}^{{\BZ}_{p}} \right]
\label{eq:projme3}
\eeq
with $P_{3} = 1 - q_{3}$, 
which gives, for the surface operator
\beq
{\CalI}_{{\Lambda}^{({\alpha})}} = 
\sum_{\left( {\lambda}^{({\alpha})} \right) \in {\rho}_{c}^{-1}({\Lambda}^{({\alpha})})} \ \ \prod_{\bnu = 0}^{p-1}
\ z_{\bnu}^{k_{\bnu -1} - k_{\bnu}} \ \cdot \ 
{\epsilon} \left[ \frac{P_{3}}{P_{1}}\ \sum\limits_{{\bnu}=1}^{p-1} \left( {\Sigma}_{\bnu +1} - {\Sigma}_{\bnu} \right) {\Sigma}_{\bnu}^{*} \right]
\eeq
which corresponds to the softly broken ${\CalN}=(4,4)$ sigma model, with the hyperk\"ahler target space $T^{*}Flags(d_{1}, d_{2}, \ldots, d_{p-1}, N)$. The mass $\ve_3$ of the adjoint hypermultiplet translates to the twisted mass corresponding to the $U(1)$ symmetry of the cotangent fibers: $t = q_{2}q_{4}$.

\subsection{The quiver construction}

Consider the $A_{2}$ quiver gauge theory, with the gauge group $U(N) \times U(N)$, 
and $N$ fundamental hypermultiplets at either node of the quiver:

\bigskip
\centerline{\includegraphics[width=5cm]{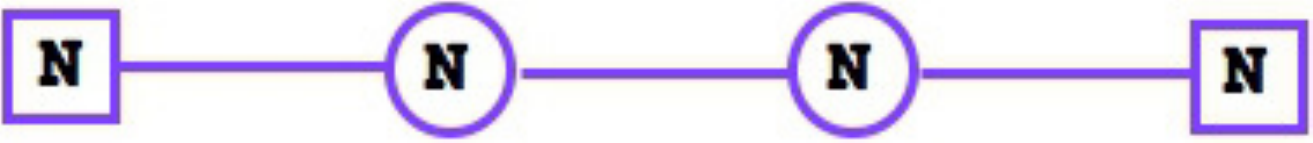}}
\bigskip

The theory depends on $2N$ masses $m_{f}^{\pm}$, $f = 1, \ldots , N$
and $2N$ Coulomb parameters ${\ac}^{\pm}_{\alpha}$, ${\alpha} = 1, \ldots , N$. 
This theory can be obtained by taking a limit ${\qe}_{0}, {\qe}_{3} \to 0$ of the ${\hat A}_{3}$-type quiver theory, which corresponds to the ${\BZ}_{4}$ orbifold of the $3,4$-space. Its partition function is given by the sum over 
$2N$-tuples of Young diagrams ${\lambda}^{ ( \pm, \alpha ) }$, $\alpha = 1, \ldots , N$. The contribution of a specific $2N$-tuple of Young diagrams is given by:
\begin{multline}
{\epsilon}\left[ \sum_{a= \pm} N_{a}K_{a}^{*} + N_{a}^{*}K_{a}q_{12} - P_{12} K_{a}K_{a}^{*} \right] \times \\
{\epsilon} \left[ - M_{+}K_{+}^{*} - N_{+}K_{-}^{*} - q_{12} K_{+}N_{-}^{*} + P_{12} K_{+}K_{-}^{*} - K_{-} M_{-}^{*} \right] 
\label{eq:instmeas}
\end{multline}
where
\beq
N_{a}^{({\tau})} = \sum_{\alpha = 1}^{N} e^{{\tau}{\ac}_{\alpha}^{a}} \, , \qquad M_{a} = \sum_{f= 1}^{N} e^{{\tau} m_{f}^{a}} 
\eeq
\beq
K_{a} = \sum_{\alpha = 1}^{N} e^{{\tau}{\ac}^{a}_{\alpha}}
\sum_{(i,j)\in {\lambda}^{(a, \alpha)}} q_{1}^{i-1} q_{2}^{j-1}
\eeq
with
$q_{1,2} = e^{{\tau}{\ve}_{1,2}}$. 

$\bullet$

Let us consider the special arrangement of masses and Coulomb parameters:
\beq
N_{+} = M_{+} \Leftrightarrow
{\ac}^{+}_{\alpha} = m^{+}_{\alpha}, \qquad {\alpha} = 1, \ldots , N
\eeq
The measure \eqref{eq:instmeas} vanishes unless ${\lambda}^{(+, {\alpha})} = {\emptyset}$ for all $\alpha$. In this case it reduces to (with $K = K_{-}$):
\beq
{\epsilon}\left[ N K^{*} + N^{*}Kq_{12} - P_{12} K K^{*}  - M_{+}K^{*}  - K M_{-}^{*} \right] 
\eeq
which is the measure of the $A_1$ theory, also known as
$U(N)$ theory with $2N$ fundamental hypermultiplets with the
masses $m^{\pm}_{f}$. 
This reduction is most simply explained using the description of the moduli space of quiver instantons \cite{N4}: the identification 
$M_{+} = N_{+}$ means that we identify the vector spaces ${\bM}_{+}$ and  ${\bN}_{+}$, thereby making the obstruction bundle ${\rm Hom}({\bM}_{+} , K_{+}) = {\rm Hom}({\bN}_{+} , K_{+})$. The latter has the equivariant section $I_{+}: {\bN}_{+} \to K_{+}$. Its zero locus is the space of the quiver ADHM data
$(B_{1,2, \pm}, I_{\pm}, J_{\pm}, \ldots )$ for which $I_{+} = 0$. 
Since (cf. \cite{N4}) $K_{+} = {\BC}[ B_{1, +} , B_{2,+} ]I_{+}({\bN}_{+})$ (in the stability chamber where all ${\zeta}^{\bR}_{a} > 0$) we get $K_{+} =0$. 

$\bullet$

Now let us impose a less stringent relation between $M_{+}$ and $N_{+}$:
\beq
M_{+} = N_{+} - P_{2} {\mu}
\eeq
where ${\mu} = e^{{\tau} ( m_{f}^{+} - {\ve}_{2} )}$ for one of the $f$'s. The equivariant section $I_{+}$ now modifies to :
\beq
B_{2,+}I_{+}{\Pi}_{f} + I_{+} ( 1 - {\Pi}_{f}) \, :\, {\bN}_{+} \longrightarrow K_{+}
\label{eq:eqsec}
\eeq
where ${\Pi}_{f} : N_{+} \to N_{+}$ is the projection onto the
eigenvector of ${\rm g}_{\rm flavor}$ with 
the eigenvalue $\mu$. 

The fixed point locus in this case allows for nontrivial $K_{+}$, however, only the row type Young diagrams are allowed:
\beq
K_{+} = {\mu} \frac{1-q_{1}^{k}}{1-q_{1}}
\eeq
for some $k \geq 0$. The instanton measure \eqref{eq:instmeas} simplifies to  
\beq
{\epsilon}\left[ {\CalN} K^{*} + {\CalN}^{*}Kq_{12} - P_{12} K K^{*}  - {\CalM}_{+}K^{*}  - K {\CalM}_{-}^{*} \right] \times {\CalI}_{f}({K})
\eeq
with $K = K_{-}$, ${\CalN} = N_{-}, {\CalM}_{+} = N_{+}$, ${\CalM}_{-} = M_{-}$ and, with $S = {\CalN} - P_{12}K$,  
\begin{multline}
{\CalI}_{f}(K) = \sum_{k=0}^{\infty} z^{k} \ {\epsilon}\left[ P_{1} {\mu} q_{1}^{k}  K_{+}^{*} + q_{12} K_{+} ({\CalM}_{+} - S )^{*} \right] = \sum_{k=0}^{\infty} \frac{k+1}{k!} \left( \frac{z}{{\ve}_{1}} \right)^{k} \ \prod_{l=1}^{k}   \frac{Y( m_{f}^{+}  + {\ve}_{1} l )}{{\tilde P}_{+} (m_{f}^{+}  + {\ve}_{1} l )}  \label{eq:ikfn}
\end{multline} 
where 
$P_{+}(x) = (x - m_{f}^{+}) {\tilde P}_{+}(x) = {\epsilon} [ - e^{{\tau}x} {\CalM}_{+}^{*}]$,
$Y(x) = {\epsilon} [ - e^{{\tau}x} S^{*}]$, and 
$z = {\qe}_{+}$ is the instanton fugacity of the $+$ node of the $A_2$ quiver, which becomes the two dimensional sigma model K\"ahler modulus.

In fact, ${\CalI}_{f}(K)$ is the $I$-function \cite{Gi1, Gi2, Gi3} (which is often confused with the $J$-function, defined using the stable maps, i.e. the nonlinear sigma model as opposed to the quasimaps, i.e. the gauged linear sigma model). There are $N$ such functions, for $f=1, \ldots , N$. Together  we should view them as the $H^{*}({\BC\BP}^{N-1})$-valued function. 

If we neglect the four dimensional instantons, i.e. replace
$Y(x)$ by the Coulomb polynomial $A(x) = \prod_{i=1}^{N} ( x - a_{i})$, the partition function \eqref{eq:ikfn} would be identified with 
the specification of \eqref{eq:grpf3} for
the ${\BC}^{N} \otimes {\CalO}(-1)$-bundle over ${\BC\BP}^{N-1}$. Here ${\BC}^{N}$ is the color space. 
In the present case, ${\ve}_{1}$ is the relevant parameter of the $\Omega$-deformation. The gauge group of
the effective sigma model is $U(1)$. The full expression 
$\CalI (K)$, which incorporates the effects of the four dimensional instantons (which generate the poles of $Y(x)$, making it the rational function of $x$) can also be interpreted
in the language of the two dimensional sigma model. However, 
the effective target space of the sigma model depends on the
four dimensional gauge field configuration. In the non-equivariant limit ${\ve}_{1}, {\ve}_{2} \to 0$, (the limit shape \cite{NO}) the operator 
${\CalI}_{f}(K)$ has the exponential asymptotics
\beq
{\CalI}_{f}(K) \sim e^{\frac{\tilde W}{{\ve}_{1}}}
\eeq
where $\tilde W$ is computed by extremizing:
\beq
{\rm log}(z) {\sigma} + \int^{\sigma} {\rm log}(Y(x)/{\tilde P}_{+}(x)) dx
\eeq
It would be nice to compare this result with the conjectures of
\cite{Gaiotto:2013sma}.  See also the further work on the surface defects and sigma models
in \cite{Bullimore:2014awa, Nawata:2014nca}, and their three dimensional lifts in \cite{Koroteev:2017nab}.

$\bullet$ One can study other types of arrangements of masses and Coulomb parameters leading to surface defects in the $A_1$ theory. For example, choose a subset $S \subset \{ 1, \ldots , N \}$, $|S| = r$, and set
\beq
M_{+} = N_{+} - P_{2} L\, , \qquad L = \sum_{s \in S} e^{{\tau}a_{+,s}} 
\eeq
Then, 
\beq
K_{+} = \sum_{s \in S} e^{{\tau}a_{+,s}} \frac{1- q_{1}^{k_{s}}}{1-q_{1}}
\eeq
 The resulting surface defect operator will correspond to the $\left[ {\rm Hom}({\CalE}, M_{+}) \to Gr(r, N) \right]$ sigma model coupled to the four dimensional gauge theory. 
The equivariant section is again given by the formula \eqref{eq:eqsec} with ${\Pi}_{f}$ now the rank $r$ projection onto the corresponding subspace of ${\bN}_{+}$. 

The surface defects of this section were also recently studied in 
  \cite{Poghossian:2016rzb},   
  \cite{Poghosyan:2016mkh}.

\subsection{The folded construction}

Finally, we can get the surface defects with the help of the folded
instanton construction \cite{N4, N5}.  Take the $U(N)$ gauge theory on the four dimensional space ${\BC}^{2}_{12}$
and the $U(M)$ gauge theory on the four dimensional space
${\BC}^{2}_{23}$. From the point of view of the observer
living on the ${\BC}^{2}_{12}$ space there are degrees of freedom propagating along the two dimensional surface ${\BC}^{1}_{2}$, which are charged in the fundamental representation. In the $\Omega$-background with the parameters ${\ve}_{1}, {\ve}_{2}, {\ve}_{3}, {\ve}_{4}$, with 
\beq
\sum_{a =1}^{4} {\ve}_{a} = 0 
\eeq
the degrees of freedom on the ${\BC}^{2}_{23}$-space can be integrated out producing the effective surface operator:
\beq
\sum_{K_{23}} \, {\qe}^{k_{23}} \, {\epsilon} \left[ - \frac{S_{23}^{*}S_{23} P_{1}}{P_{23}^{*}} - 
\frac{S_{12} S_{23}^{*} q_{23} + S_{12}^{*}S_{23} q_{12}}{P_{2}} \right]
\label{eq:fold}
\eeq
If we freeze the bulk dynamics both in the ${\BC}^{2}_{12}$
and the ${\BC}^{2}_{23}$ spaces, the interaction \eqref{eq:fold}
corresponds to the two dimensional ${\CalN}=(2,2)$ theory with 
two chiral bi-fundamental multiplets, one in the $(N, {\bar M})$
another in the $({\bar N}, M)$ representation of $U(N) \times U(M)$, with the twisted masses given by ${\ve}_{1} + {\ve}_{2}$
and ${\ve}_{2} + {\ve}_{3}$, respectively. 

One can also add the orbifold defects on top of the folded construction, with $(z_{2}, z_{4}) \mapsto ({\varpi}z_{2}, {\varpi}^{-1}z_{4})$. Then, effectively, we are getting a junction of two surface effects, one extended along the ${\BC}^{1}_{1}$ plane, another along the ${\BC}^{1}_{3}$ plane. This is a singular rational curve which can be deformed to the smooth curve. However, the deformation parameter has the equivariant weight ${\ve}_{1} + {\ve}_{3}$.

In the companion paper \cite{N6} will analyze these surface defects using the $qq$-characters. They are obtained by adding another component, e.g. ${\BC}^{2}_{34}$, or ${\BC}^{2}_{14}$, to the worldvolume of the gauge theory.

The compactness theorem of \cite{N4} is powerful
when ${\rm Hom}_{{\BZ}_{p}} (N_{12} , N_{34}\otimes {\CalR}_{1}) \approx {\BC}$
and
${\rm Hom}_{{\BZ}_{p}} (N_{34} \otimes {\CalR}_{1} , N_{23}) = 0$ for ${\BC}^{2}_{34}$ $qq$-operator, and, analogously
${\rm Hom}_{{\BZ}_{p}} (N_{12} , N_{14}\otimes {\CalR}_{1}) = 0$
and
${\rm Hom}_{{\BZ}_{p}} (N_{14} \otimes {\CalR}_{1} , N_{23}) \approx {\BC}$ for the ${\BC}^{2}_{14}$ operator. This is easy to arrange, e.g. for $p = N+M$. We shall study the corresponding KZ-like equations in \cite{N6}. 

\section{$qq$-characters in two dimensional sigma models}
 
In \cite{N2, N3, N4} we introduced the so-called $qq$-character
observables in four dimensional ${\CalN}=2$ theories. They are convenient to express the non-perturbative Dyson-Schwinger
equations. 

In the two dimensional context one also expects the non-perturbative Dyson-Schwinger equations to take place and to be useful. We shall be working in the softly broken ${\CalN}=(4,4)$ quiver gauge theory case. 

Define, for $i \in \Vg$, 
\beq
Y_{i}(x) = \frac{Q_{i}(x)}{Q_{i}(x-2u)}\, , \qquad D_{i}(x) = \frac{P_{i}(x + u)}{P_{i}(x-u)}
\label{eq:ydp}
\eeq
We shall view $Q_{i}(x)$ and $Y_{i}(x)$ as the vacuum expectation values of the observables
${\bQ}_{i}(x)$ and ${\bY}_{i}(x)$:
\beq
{\bQ}_{i}(x) = {\rm Det}_{{\bN}_{i}} \left( x - {\sigma}_{i} \right) \, , \qquad {\bY}_{i}(x) = \frac{{\bQ}_{i}(x)}{{\bQ}_{i}(x-2u)}
\label{eq:qyobs}
\eeq
Then \eqref{eq:bae} reads:
\beq
 Y_{i}({\xi}+2u) \ = \  - \, {\tilde\qe}_{i} \  \frac{D_{i}({\xi})}{Y_{i}({\xi})}\, \prod_{e \in t^{-1}(i)} \,Y_{s(e)}({\xi}+u+m_{e})
\prod_{e \in s^{-1}(i)}  \,Y_{t(e)}({\xi}+u-m_{e})
\label{eq:cancelpoles}
\eeq
for all ${\xi}$, s.t. $Q_{i}({\xi}) = 0$, with
\beq
{\tilde\qe}_{i} =  {\qe}_{i} \, (-1)^{m_{i} + n_{i}-1} \prod_{j \in \Vg} (-1)^{C_{ij} n_{j}}
\eeq
The relations \eqref{eq:cancelpoles} can be interpreted as the reflections
generating the generalization of the Weyl group ${\CalW}_{\Gamma}$ \cite{NP1,NP2}. The natural step is to introduce the Weyl-invariant, which is characterized by the property that the only poles in $x$ is has come from the poles of $P_{i}$'s, i.e. all the superficial poles coming from $Q_i$'s cancel between themselves. Thus, for every $i \in \Vg$ we define:
\beq
{\bf T}_{i}(x) = {\bY}_{i}(x+2u) + {\tilde\qe}_{i} \ \frac{D_{i}(x)}{{\bY}_{i}(x)}
\prod_{e \in t^{-1}(i)} \,{\bY}_{s(e)}(x+u+{\mu}_{e})
\prod_{e \in s^{-1}(i)}  \,{\bY}_{t(e)}(x+u-{\mu}_{e}) + \ldots 
\label{eq:qqchar}
\eeq 
As we recalled above, for affine $ADE$ quivers $\gamma$ the two dimensional quiver gauge theory can be engineered as the low energy limit of the theory on a stack of fractional $D1$ strings \cite{DM} localized at the tip $0$ of the $ADE$ orbifold singularity $0 \in {\BC}^{2}/{\Gamma}_{\gamma}$. 
As we explained in \cite{N3,N4,N5, NP} the observable \eqref{eq:qqchar} is the result of integrating out the open strings, connecting these branes and the stack of (fractional)
 $D3$-brane wrapping the ALE space ${\BC}^{2}/{\Gamma}_{\gamma}$.

The full expression for the $qq$-character \eqref{eq:qqchar}
is identical to that given in \cite{N3}, the sum of integrals over the Nakajima varieties corresponding to the quiver $\gamma$. 
It makes sense for all quivers, not just the ADE type. 
The difference between the two dimensional and the four dimensional cases is the nature of the ${\bY}_{i}(x)$-observables and the $D_{i}(x)$-operators. In two dimensions both ${\bY}_i(x)$ and $D_i (x)$ are degree zero ratios of polynomials, as in \eqref{eq:ydp}. In four dimensions \cite{N3}
$D_i(x)$ are the polynomials whose roots are the masses of the
fundamental hypermultiplets, while $Y_{i}(x) = {\epsilon} [ - e^{x} S_{i}^{*} ]$ is the value of ${\bY}_{i}(x)$ in the instanton background, it is a degree $n_i$ rational function of $x$. 

The $qq$-character is a local observable in both two and four dimensional theories. By exploiting the analytic properties of
its expectation values one can derive numerous Ward identities. 

In the limit $\ve \to 0$ the $qq$-character becomes the $q$-character (more precisely, its Yangian version). The algebraic theory of $q$-characters was introduced in \cite{FR}. Their physical use was found first in \cite{NP2} in the applications
to the four and five dimensional ${\CalN}=2$ theories subject
to the two-dimensional $\Omega$-deformation. 

\section{Conclusions and outlook}

Here are some topics we have left out. 

We discussed two dimensional supersymmetric gauge theories
coupled to the four dimensional ones. In the quiver construction
the two dimensional theory arises as the effective description of the four dimensional theory, similar to the cosmic string theory to which the ${\CalN}=2$ theory reduces with the special mass deformation which vanishes along a complex curve \cite{Witten:1994ev}. Here the role of the mass deformation is played by the Higgsing of the gauge group (more precisely the gauge group factor) in the presence of $\Omega$-deformation. It would be nice to have a more detailed field theoretic description of this Higgsing and the generation of the mass gap which vanishes along a codimension two surface. Perhaps the
noncommutative description of the section $\bf 5$ could be used in this regard. See also \cite{Shifman:2014lba} for the similar string solutions. 

One should note that there is a simpler version of this construction, where the whole four dimensional theory becomes effectively the two dimensional one. Of course, any four dimensional theory can be viewed as a two dimensional theory with an infinite set of fields. Here we talk about the
equivalence of the four dimensional theory and conventional two dimensional field theory. See \cite{Chen:2011sj, Dorey:2011pa, Dorey:2015gfn} for numerous examples. 

The $qq$-characters which we discussed in the section $\bf 6$
can be used to give a two dimensional gauge theory realization
of the integrable lattice systems built using the $R$-matrices of the Yangians (or quantum affine algebras). This realization is used in the explanations\footnote{See the author's contribution to Strings-Math'2017 in Hamburg} of the duality relating the Bethe/gauge correspondence and the recently revived four dimensional version of Chern-Simons theory \cite{AFM, GN3, NPhD, C1, C2, W4}.

The moduli space of vortex solutions in the two dimensional theory on the surface of defect coupled to the four dimensional instantons can be given an ADHM-like description, by taking the
quiver instanton moduli space \cite{N4} and imposing the
additional equations \eqref{eq:eqsec}. It would be interesting to compare the resulting ADHM-like equations to those in 
\cite{Bruzzo:2010fk}. Also, our moduli spaces require the four dimensional ADHM-like data. The (Grassmanian) sigma models can be described by fewer degrees of freedom. The connection between these two description is, perhaps, the manifestation of the renormalization group flow argued in \cite{Frenkel:2015rda}.

There is an interesting class of local observables in supersymmetric gauge theory, which are obtained by lifting the (partially) twisted theory on the blowup ${\pi}: {\hat X}_{p} \to X$ of the spacetime at a point $p$, 
and integrating out the degrees of freedom associated
with the exceptional divisor ${\pi}^{-1}(p)$. Using the quiver
description of the moduli space of instantons on the blowup  \cite{K, NY, NY1} one can express the resulting operator as
a series in the ${\bY}$-observables, similar to \eqref{eq:ikfn}. 
Also, by comparing the $\Omega$-deformed theories with the surface defects on the blowup to the original theory one can derive a set of Hirota-like relations involving the surface defect
partition function and the bulk partition function. These relations explain the results of \cite{Gamayun:2012ma}.

It would be nice if the folded construction could be used to derive the differential/difference equations obeyed by the partition functions \eqref{eq:instpf}, connecting to the results of \cite{Maulik:2012wi}, \cite{Okounkov:2016sya} (see \cite{Koroteev:2018azn} where such equations are derived, for the $A$-type quivers, using the methods of \cite{Maulik:2012wi}). 

The orbifold defects we studied in the four dimensional gauge theory can also be defined in the two dimensional gauged linear sigma models. In certain cases these produce the wavefunctions of the Bethe states of the quantum integrable systems dual to the ${\CalN}=(2,2)$ theory via Bethe/gauge correspondence \cite{NBd, Bullimore:2017lwu, Gorsky:2017hro}. However, these constructions require the theory to have a large flavor symmetry. It would be nice to understand
the results of \cite{Aganagic:2017gsx} in these terms. More generally, 
the stable envelope basis of \cite{Maulik:2012wi} should translate to the canonical basis in the space of supersymmetric codimension two defects. 

We mostly talked about the two dimensional defects within four dimensional theories. Everything we said, including the $qq$-characters (hence the name) lifts to the three-dimensional defects within five dimensional theories, compactified on a circle of radius $R$ (see, e.g. \cite{Koroteev:2017nab}). One can get new observables in four dimensional theories not only by sending the radius $R$ back to zero, but also by considering the extreme limits of the orbifold defects we discussed. 
In particular, the $p \to \infty$ limit of the ${\BZ}_{p}$ orbifold defect in the theory on ${\BR}^{4} \times {\bS}^1_{R}$ becomes 
a supersymmetric boundary condition in the four dimensional theory on ${\BR}^{2} \times {\BR}^1_{+} \times {\bS}^{1}_{R}$. 
Another useful construction involves the blowup-type defects modeled on the ALE singularity ${\BC}^{2}/{\BZ}_{p}$. In the limit $p \to \infty$ the theory on ${\BC}^{2}/{\BZ}_{p} \times {\bS}^{1}_{R}$ becomes the theory on ${\BR}^{3} \times {\bS}^{1}_{R}$ together with the monopole-like singularity at the origin in ${\BR}^{3} = {\BC}^{2}/U(1)$, corresponding to the 't Hooft line operator wrapping ${\bS}^1_{R}$. In this way we get, essentially for free, the localization computation of the vacuum expectation values of the 't Hooft (and more general dyonic) operators in the ${\CalN}=2$ theory compactified on a circle. The computations in \cite{Ito:2011ea, Gomis:2011pf, Chang:2015ofn} confirm this. In this way one can hope to bring the conjectures of \cite{GMN} within the reach of the conventional proofs.

There are many more topics to explore and elucidate.

\section{\textbf{Acknowledgements}.}\ Research was partly supported
by the National Science Foundation under grant
no.~NSF-PHY/1404446. Any opinions, findings, and conclusions or
recommendations expressed in this material are those of the authors
and do not necessarily reflect the views of the National Science
Foundation. 

{}The author thanks A.~Losev for discussions about two dimensional topological theories in 1992-1994,  E.~Frenkel for the invitation to the DARPA program `Langlands Program and Physics' conference at the IAS in March 8-10 2004, and for the opportunity to present there some of the ideas developed in this paper, as well as for numerous patient explanations on \cite{FR} and other topics. The author is grateful to A.~Okounkov, V.~Pestun and A.~Rosly for numerous useful discussions concerning the topics of this paper over the recent years, as well as to S.~Jeong and O.~Tsymbaliuk for the collaboration on the related projects. 

{}The paper was finished while the author visited the IHES (Bures-sur-Yvette). We thank this remarkable institution for its hospitality.

\section{Appendix A. Defects in quantum mechanics}

Take two quantum systems, $({\hat H}_{1}, {\CalH}_{1})$ and $({\hat H}_{2}, {\CalH}_{2})$ with the Hilbert spaces ${\CalH}_{i}$ and the Hamiltonians ${\hat H}_{i}$, $i  = 1,2$.  The spacetime manifolds $X_{1}$ and $X_{2}$ are, say, two intervals $X_{2} = [0, t_{2} ]$, $X_{1} = [ - t_{1}, 0 ]$, with $t_{1,2} > 0$, the intersection $X_{12}$ consist of one point, $t=0$. 

\bigskip

\centerline{\includegraphics[width=2cm]{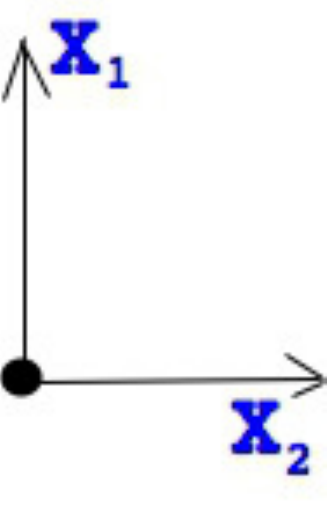}}

\bigskip

Then the path integral of the combined system computes:
\beq
\langle {\psi}_{2} \vert_{{\CalH}_{2}}\ e^{-{\ii}t_{2} {\hat H}_{2}} \ {\Pi}_{12} \ e^{-{\ii}t_{1} {\hat H}_{1}} \ \vert {\psi}_{1} \rangle_{{\CalH}_{1}}
\label{eq:h1h2}
\eeq
where ${\Pi}_{12} : {\CalH}_{1} \longrightarrow {\CalH}_{2}$ is an operator acting between the two Hilbert spaces, $\vert {\psi}_{1} \rangle_{{\CalH}_{1}} \in {\CalH}_{1}$, $\vert {\psi}_{2} \rangle_{{\CalH}_{2}}$ are the states of two systems, representing the boundary conditions at the other end of each interval. In the coordinate representation, the operator ${\Pi}_{12}$ has the kernel:
\beq
{\Pi}_{12} ({\phi}_{1}, {\phi}_{2}) = \int\, [D{\bf\Psi}] \, e^{-L_{12}({\bf\Psi} ; {\phi}_{1}, {\phi}_{2})}
\eeq
where ${\phi}_{i} = {\bf\Phi}_{i}\vert_{X_{12}}$, $i=1,2$. 

To make the example even more specific, although somewhat artificial, let us take  the space ${\CalH}_{1} = {\BC}[ a_{1}^{\dagger}, \ldots , a_{p}^{\dagger} ]\, \vert \, {\rm vac} \, \rangle$ of states of $p$ bosonic oscillators $[a_{i}, a_{j}^{\dagger}] = {\delta}_{ij}$, with
the pseudovacuum $\vert \, {\rm vac} \, \rangle$ annihilated by $a_{i}$'s, with the generic quadratic Hamiltonian
\beq
{\hat H}_{1} = \sum_{i,j=1}^{p} \, K_{ij} a_{i}^{\dagger}a_{j} + B_{ij} a_{i}a_{j} + {\bar B}_{ij} a_{i}^{\dagger} a_{j}^{\dagger}
\eeq
with some matrices $K, B$ ($B$ sufficiently small), 
and ${\CalH}_{2} \approx {\BC}^{N}$, $N = 2^{L}$,  
the space of states of $L$ fermionic oscillators $\{ c_{A}, c_{B}^{\dagger} \} = {\delta}_{AB}$, with the Hamiltonian, say
\beq
{\hat H}_{2} = \sum_{A, B=1}^{N}\, M_{AB} c_{A}^{\dagger} c_{B} + N_{AB} c_{A}c_{B} + {\bar N}_{BA} c_{A}^{\dagger} c_{B}^{\dagger}
\eeq
with some matrices $M, N$. 

The operator $P_{12}$ orthogonally projects the space ${\CalH}_{1}$ 
onto the subspace of ${\hat Q}^{B}$ charge 
\beq
q = L +1 - p
\eeq
states, which is then isomorphically mapped to  the charge ${\hat Q}^{F} = q$ subspace of ${\CalH}_{2}$. Here the charges are defined  by
\beq
{\hat Q}^{B} = \sum_{i=1}^{p} a_{i}^{\dagger}a_{i}, \qquad {\hat Q}^{F} = \sum_{A=1}^{L} c_{A}^{\dagger}c_{A}
\eeq
Explicitly:
\beq
{\Pi}_{12}\left( \prod_{i=1}^{p} \frac{a_{i}^{n_{i}}}{\sqrt{n_{i}!}} \vert \, {\rm vac} \, \rangle \right) \ = \ \begin{cases}
\quad c_{n_{1}+1}^{\dagger} c_{n_{1}+n_{2}+2}^{\dagger} \ldots c_{n_{1}+\ldots +n_{p-1} + p-1}^{\dagger}\ \vert {\Omega}_{F} \rangle \,  , \qquad\qquad\qquad\qquad n_{1} + \ldots + n_{p} = q \\
\\
\quad 0 \, , \qquad  n_{1} + \ldots + n_{p} \neq q \end{cases}
\eeq
where the pseudovacuum state $\vert {\Omega}_{F} \rangle$ is annihilated by all $c_{A}$'s.

\section{Appendix B. More on Yang-Mills theory on punctured surfaces}

Let us discuss the analogue of the Yang-Mills connections in the parabolic setting. 
We want to extremize the Yang-Mills action together with the action of the sources coming from the Wilson points:
\beq
-S_{\rm ym} = {\ii} \int_{\Sigma_{g}} {\Tr} ({\sigma}F_{A}) + \frac{g_{\rm ym}^{2}}{2} \int_{\Sigma} {\Tr} {\sigma}^{2} + {\ii} \sum_{i=1}^{n} {\Tr} {\sigma}(x_{i}) J_{i} 
\label{eq:yma}
\eeq
so that the path integral measure was $e^{-S_{\rm ym}}$, where $J_{i} \in O_{{\nu}_{i}}$ is allowed to vary as well. Varying w.r.t $J_{i}$ we get: $J_{i}$ is one of the fixed points of the $G$-action on $O_{{\nu}_{i}}$ generated by ${\sigma}(x_{i})$, while
\beq
\begin{aligned}
&  F_{A} + \sum_{i} J_{i} {\delta}^{(2)}(x_{i}) =  {\ii} g_{\rm ym}^{2} {\sigma} \\
& d_{A}{\sigma} = 0 \\
\end{aligned}
\label{eq:ymp}
\eeq
the equations, for generic non-zero $\sigma$, imply that the structure group $G$ of the bundle reduces to $T$, its maximal torus, ${\sigma}, A, J_{i} \in {\mathfrak t}$, moreover: 
${\sigma} = {\rm const}$, $F_{A} = f  + da$, $a \in {\Omega}^{1}({\Sigma}_{g}) \otimes {\mathfrak t}$ and $f$ is a harmonic two-form on $\Sigma_g$ valued in $\mathfrak t$, moreover
\beq
\int_{\Sigma_g} f \in 2\pi  {\Lambda}_{\rm cw} = {\rm Hom} (U(1), G)
\eeq
Now, \eqref{eq:ymp} implies, for abelian $A, f, a, J_i$
\beq
\int_{\Sigma_{g}} f  + \sum_{i=1}^{n} J_{i} = - {\ii}g_{\rm ym}^{2} A_{\Sigma_{g}} {\sigma}
\eeq
which means that $\sigma$ takes quantized values as in the unramified case, but the lattice is shifted by the amount which depends on $J_i$'s. Moreover, the choices of $J_i$'s are labelled by the cosets $W/W_{{\nu}_{i}}$, $i=1, \ldots , n$, so we get a multitude of lattices labelled by
\beq
\varprod\limits_{i=1}^{n} \, W/W_{{\nu}_{i}}
\eeq
Moreover, the overall $W$-action is a symmetry, so finally, the set of critical points is the quotient
\beq
\left( {\Lambda}_{\rm cw} \times \varprod\limits_{i=1}^{n} \, W/W_{{\nu}_{i}} \right) \, {\slash}\, W
\eeq
Let us make an explicit calculation  for $G = SU(2)$. The orbits $O_{{\nu}_{i}} \approx {\BC\BP}^{1}$ for non-zero ${\nu}_{i}$'s, and $W = {\BZ}_{2}$. 
Let us assume the eigenvalues $\pm {\ii}{\nu}_{i}$ of the traceless two by two antihermitian matrices $J_{i}$, $i = 1, \ldots, n$ are in generic position. Let the metric on ${\Sigma}_{g}$ be
\beq
ds^2 = {\rm g}_{z\zb} dzd{\zb}, \qquad \frac{1}{2\ii} \int_{{\Sigma}_{g}} {\rm g}_{z\zb} dz \wedge d{\zb} = A_{{\Sigma}_{g}}
\eeq
Then, the generalized Yang-Mills connection has $f =  m \frac{\pi  }{A_{{\Sigma}_{g}}} {\rm g}_{z\zb} dz \wedge d{\zb}$, $m \in {\BZ}$, 
\[  {\sigma} = - \frac{1}{g_{\rm ym}^{2} A_{{\Sigma}_{g}}} \left(  2\pi m + \sum_{i=1}^{n} \pm {\nu}_{i} \right) \] and $a = {\ii} \star d{\varphi}$ with 
\[ {\Delta}{\varphi}  = - \sum_{i=1}^{n} \pm {\nu}_{i} \left( {\delta}^{(2)}(z-x_{i}) - \frac{1}{A_{{\Sigma}_{g}}} \right)   \] 
Despite the appearences, there is no singularity in the value of the action:
\[ -S_{\rm ym} = - g_{\rm ym}^2 A_{S^2} {\sigma}^2 = - \frac{4{\pi}^2}{g_{\rm ym}^2 A_{{\Sigma}_{g}}} \left(   m + \sum_{i=1}^{n} \pm \frac{{\nu}_{i}}{2\pi}  \right)^2  \] 
Cf. the partition function:
\beq
Z (g_{\rm ym}^2 A_{\Sigma_{g}}; {\bnu}) = \frac{1}{(2{\pi}{\ii})^{n}}
\sum_{{\tilde m} \in {\BZ}} \, 
\frac{e^{- g_{\rm ym}^2 A_{\Sigma_{g}} {\tilde m}^2}}{{\tilde m}^{n-2+2g}}\,  \prod_{i=1}^{n} \, \left(
e^{{\ii}{\nu}_{i}{\tilde m}} - e^{-{\ii}{\nu}_{i}{\tilde m}} \right) \
= \ - \frac{1}{{\pi}^2} \sum_{e \in {\bf 2}^{n-1}} (-1)^{e} {\gamma}_{n-2+2g} \left( {\bnu}_{e} ; g_{\rm ym}^2 A_{\Sigma_{g}}\right) 
\label{eq:ympf}
\eeq
where ${\bf 2}^{n-1}$ is the set of $n$-tuples of $\pm 1$, up to the overall change 
of sign, 
${\bnu}_{e} = \sum_{i=1}^{n} \pm {\nu}_{i}$, 
$(-1)^{e} = \prod_{i=1}^{n} \pm 1$, 
\[ {\gamma}_{k}(x ; t) = \begin{cases} \qquad
\sum\limits_{{\tilde m} =1}^{\infty}
\frac{{\rm cos}({\tilde m}x) }{(2{\pi}{\ii}{\tilde m})^{k}} \, e^{- t {\tilde m}^{2}} \,  , &  k - {\rm even} \\
& \\ \qquad
 \sum\limits_{{\tilde m} =1}^{\infty}
\frac{{\rm sin}({\tilde m}x) }{(2{\pi}{\ii}{\tilde m})^{k}} \, e^{- t {\tilde m}^{2}} \,   , &  k - {\rm odd} \end{cases} \]
Now, 
\beq
(4{\pi}^2 {\pa}_{t})^{l} {\gamma}_{2l} (x; t) = - \frac 12  + \frac 12 \sum\limits_{{\tilde m} \in {\BZ}} e^{{\ii}{\tilde m}x  - t{\tilde m}^{2}}  = -\frac 12 + \frac{1}{\sqrt{8{\pi}t}} \sum_{n\in {\BZ}}
e^{-\frac{(x+ 2{\pi}n)^2}{4t}} 
\eeq
which means that as long as $x \notin 2{\pi}{\BZ}$, ${\gamma}_{2l}(x; t)$, for small
$t >0$ behaves as a polynomial of degree $t$ plus exponentially small corrections. Similarly
\beq
(2{\pi}{\ii}{\pa}_{x}) (4{\pi}^2 {\pa}_{t})^{l} {\gamma}_{2l+1} (x; t) = -\frac 12 + \frac{1}{\sqrt{8{\pi}t}} \sum_{n\in {\BZ}}
e^{-\frac{(x+ 2{\pi}n)^2}{4t}} 
\eeq
Finally, the heat equation 
\beq
\left( (2{\pi}{\ii}{\pa}_{x})^{2} + {\pa}_{t}  \right) {\gamma}_{k}(x;t) = 0
\eeq
relates the $t$ and the $x$ expansion for small $t$ and $x$.

The small $g^2A_{S^2}$ expansion of \eqref{eq:ympf} can be done using the 
following elementary function:
\begin{multline}
\sum_{{\tilde m}=1}^{\infty} \frac{{\rm cos}({\tilde m}{\theta})}{{\tilde m}^{2}} = 
\frac{\pi^2}{6} - \frac{{\theta}(2{\pi} - {\theta})}{4}\, , \qquad {\rm for} \qquad
0 \leq {\theta} \leq 2\pi \\
= - \frac{{\pi}^{2}}{12} - {\rm min}_{n \in {\BZ}} \ \frac{({\theta} - (2n+1){\pi})^2}{4}
\end{multline}
which gives, for $n=4$:
\begin{multline}
Z (g^2 A_{S^2}; {\nu}_{1}, {\nu}_{2}, {\nu}_{3}, {\nu}_{4}) = \\
\frac{1}{(2{\pi}{\ii})^{n}}
\sum_{{\tilde m} \in {\BZ}} \, 
\frac{e^{- g^2 A_{S^2} {\tilde m}^2}}{{\tilde m}^{2}}\,  \prod_{i=1}^{n} \left(
e^{{\ii}{\nu}_{i}{\tilde m}} - e^{-{\ii}{\nu}_{i}{\tilde m}} \right) 
= - \frac{1}{2{\pi}^2} \sum_{e \in 2^{n}} (-1)^{e} {\gamma}_{n-2} \left( {\bnu}_{e} ; g^2 A_{S^2}\right) 
\label{eq:4ptpf}
\end{multline}
The piece-wise polynomial dependence on ${\nu}_{i}$'s is a consequence of the Duistermaat-Heckman's theorem \cite{DH}. Indeed, we can view $\nu_{i}$'s as the levels of the abelian moment maps by writing
\beq
J_{i} = I_{i}I_{i}^{\dagger} - \frac 12  \, I_{i}^{\dagger} I_{i} \, {\bf 1}_{2}
\eeq 
with 
\beq
I_{i}^{\dagger}I_{i} = {\nu}_{i} 
\eeq
being the $U(1)_{i}$ moment map.

\end{document}